\documentclass[aps,pre,reprint,superscriptaddress,nofootinbib,floatfix]{revtex4-2}
\usepackage{graphicx}
\usepackage{amsmath}
\usepackage{bm}
\usepackage{subcaption}
\usepackage{hyperref}
\usepackage{cleveref}
\usepackage{color}
\newif\ifdark
\ifdark
  \pagecolor{black}
  \makeatletter
  
  \AtBeginDocument{%
    \color{white}%
    \g@addto@macro\frontmatter@title@format{\color{white}}%
    \g@addto@macro\frontmatter@authorformat{\color{white}}%
    \g@addto@macro\frontmatter@affiliationfont{\color{white}}%
    \g@addto@macro\frontmatter@abstractfont{\color{white}}%
    \g@addto@macro\frontmatter@RRAP@format{\color{white}}%
    \let\rvtx@orig@maketitle\maketitle
    \def\maketitle{\rvtx@orig@maketitle\color{white}}%
  }
  \makeatother
\fi

\begin{document}

\title{Enhanced alpha channeling with spin-polarized fuel}

\author{J. F. Parisi}
\affiliation{Marathon Fusion, 150 Mississippi Street, San Francisco, CA, 94107, USA}
\author{ A. Diallo}
\affiliation{Princeton Plasma Physics Laboratory, 100 Stellarator Road, Princeton, NJ 08543, USA}
\affiliation{Proxima Fusion GmbH, Fl\"o\ss{}ergasse 2, Munich, 81369, Germany}
\author{ J. W. S. Cook}
\affiliation{United Kingdom Atomic Energy Authority, Culham Campus, Abingdon, Oxfordshire, OX14 3DB, UK}
\affiliation{Centre for Fusion, Space and Astrophysics, Department of Physics, University of Warwick, Coventry, CV4 7AL, UK}

\begin{abstract}
The nuclear spin state of deuterium-tritium (D-T) fuel sets both the D-T fusion cross section and the emission direction distribution of fusion-born alphas and neutrons. We show two ways that spin-polarized fuel (SPF) could enhance alpha channeling, the wave-mediated damping of alpha power onto fuel ions rather than electrons, which is predicted to increase fusion power significantly. First, the enhanced SPF cross section produces more alphas, and second, the SPF perpendicular (to the magnetic field) bias of the alphas' kinetic energy heats ions more efficiently through a perpendicular-resonant wave. The birth anisotropy survives slowing-down and appears as a population inversion of the bulk alpha distribution $f$, with $\partial f/\partial v_\perp>0$ over a broad region, where $v_\perp$ is the velocity perpendicular to the magnetic field, so resonant alphas can drive a suitably tuned channeling wave rather than damp it. A transport calculation of SPF without alpha channeling roughly doubles the fusion power density from the cross-section boost and its temperature feedback on the reactivity, a well-known result. Our velocity-space calculations find the channeling efficiency about 1.5 times higher for the perpendicular alpha distribution of vector-aligned fuel than for unpolarized fuel, and channeling raises the fusion power enhancement to three or four times as the channeling efficiency improves, provided the waves do not depolarize the fuel. A transport model of an ARC-class equilibrium with stiff critical-gradient transport gives a fusion power enhancement of $2.2$, rising to $3.4$ for less stiff transport. Letting the critical gradients rise with the hotter ions raises the enhancement to $4.7$ in the zero-dimensional model. Channeling also transports helium quickly to the divertor, lowering the core helium fraction relative to the exhaust: at fixed divertor pumping the core helium fraction nearly halves, and a divertor pump several times less selective for helium supports the same core helium dilution. Spin-polarized fuel thus enhances fusion power through the anisotropic alpha distribution, beyond its increase of the reactivity.
\end{abstract}

\maketitle

\section{Introduction}\label{sec:intro}

A burning deuterium-tritium (D-T) fusion plasma is mainly heated by alpha particles produced in fusion reactions. This self-heating is valuable because it reduces the power and capital costs of external systems that heat the plasma. However, just as important as the magnitude or fraction of alpha heating power~\cite{Lawson1957,Wurzel2022} is \emph{how} that alpha power is partitioned into the plasma's ions and electrons. It is generally preferable for the alphas to directly heat ions, rather than the standard scenario of alphas mainly heating the electrons. Direct heating of fuel ions by alphas mediated by waves is known as alpha channeling~\cite{fisch1992,fisch1995alpha}. 

In this work we combine two techniques, spin-polarized fuel~\cite{kulsrud1982} and alpha channeling, and show that they can positively reinforce each other. Spin polarization makes more alphas and emits them perpendicular to the magnetic field, the population that the perpendicular-resonant waves used for channeling extract best: the alpha power then heats the fuel ions directly, and the wave sweeps the helium ash out of the core as it drains each alpha. Together these raise the fusion power density while lowering the tritium inventory, the helium pumping requirement, and the auxiliary heating demand, each a leading cost driver for a fusion power plant (FPP)~\cite{Wade2021}.

A magnetically confined D-T plasma reaches high fusion gain and higher power density more easily when fusion-born alpha particles return more of their energy to the fuel ions, since the reactivity is ion temperature-dependent, the electrons radiate, and at fixed plasma pressure the fusion output is maximized by holding the ions hotter than the electrons~\cite{fisch1992}. Without intervention, an alpha born at $\sim3.5$ MeV slows down mainly on the electrons, heating them rather than the fuel ions and leaving cooled helium ash behind in the fusion core. Alpha channeling achieves multiple goals at once, damping the alpha energy on the fuel ions instead of the electrons while moving the helium ash radially outward for exhaust~\cite{fisch1992,fisch2015}. When an alpha exchanges energy with a wave of frequency $\omega$ and toroidal mode number $n_\phi$, the alpha's toroidal angular momentum $P_\phi$ changes with its energy $E$ according to $dP_\phi/dE=n_\phi/\omega$. Because $P_\phi\approx-Ze\,\psi(r)$ apart from a small orbit correction, with $Ze$ the alpha charge and $\psi$ the poloidal flux rising with minor radius $r$, a change in $P_\phi$ is a change in radial location. Choosing $\omega$ and $n_\phi$ so that these paths connect the energetic alphas in the core to low-energy helium-4 in the plasma edge, the alphas diffuse down the paths and give their energy to the wave, which damps on the fuel ions.

\begin{figure*}[tbp]
\includegraphics[width=0.92\textwidth]{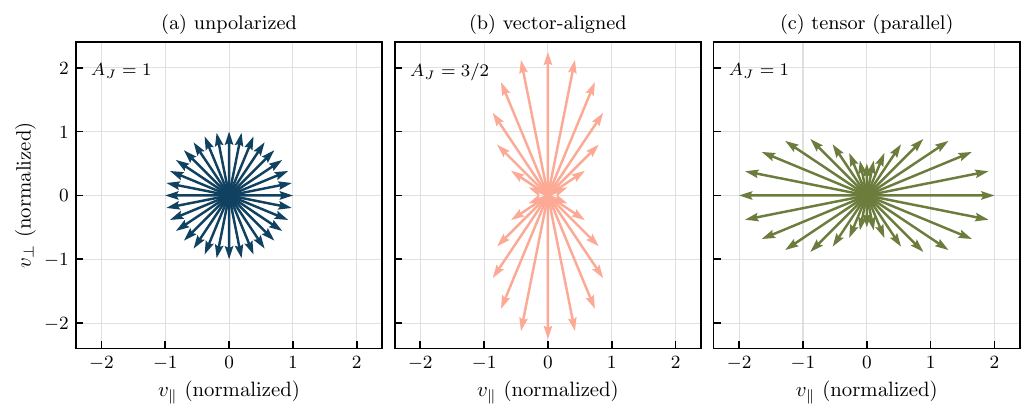}
\caption{Birth emission probability (proportional to arrow length) of fusion alphas from the D-T fusion reaction site for three spin-polarized configurations, with cross-section enhancement $A_J$.}
\label{fig:emission}
\end{figure*}

The waves most studied for channeling, ion Bernstein and lower-hybrid waves, have perpendicular wavenumber far exceeding parallel, $k_\perp\gg k_\parallel$, and remove energy from the alphas' perpendicular motion~\cite{fisch2015}, so a population with more perpendicular than parallel energy couples to them most strongly. If the alphas amplify the wave rather than damp it, the same excess of perpendicular energy sets how fast the wave grows~\cite{cianfrani2018,cookdendy2017}. 

Alpha channeling has been developed extensively in theory and computation, including calculations of the extractable power~\cite{fischherrmann1994}, the excitation of the required wave spectra~\cite{valeofisch1994}, two-wave schemes that remove perpendicular and parallel energy separately~\cite{herrmann1997}, momentum-conserving treatments~\cite{ochsfisch2021}, tokamak implementations with lower hybrid waves~\cite{ochsbertelli2015a,ochsbertelli2015b}, rotating-plasma variants~\cite{fetterman2008}, reversed magnetic shear~\cite{wei2022core}, and advanced fuels~\cite{ochs2022,kolmes2022wave}. Initial experimental support came from TFTR, where mode-converted ion Bernstein waves interacted with energetic ions standing in for alphas, in D-${}^{3}$He~\cite{fisch1997tftr} and in D-T discharges~\cite{darrow1996}, heating D-D fusion tritons or accelerated deuterium neutral beam ions above their birth energy and moving them across the passing-trapped boundary as channeling predicts. An experimental demonstration of channeling from fusion-born alphas remains to be shown.

Another technique to increase the D-T fusion reactivity is polarizing the nuclear spin of D and T fuel ions. Aligning the nuclear spins of deuterium and tritium parallel to the magnetic field $\bm{B}$ makes the D-T fusion reaction more likely, raising its cross section by a factor $A_J\le3/2$~\cite{kulsrud1982,kulsrud1986,hupin2019}. A second consequence is that the fusion products are emitted anisotropically about $\bm{B}$~\cite{kulsrud1982,kulsrud1986}. \Cref{fig:emission} shows the emission patterns for the three polarization modes considered in this work: isotropic emission for unpolarized fuel (a), perpendicular-biased for vector-aligned fuel (b), and parallel-biased for tensor-polarized fuel (c).

This anisotropy has been viewed either as a problem, because it can drive ion-cyclotron emission and depolarization~\cite{coppi1983,cowley1986,cook2026}, or as a useful way to modify the $14$ MeV neutron load in the surrounding blanket~\cite{bae2025,bruhaug2024,schwartz2025}. Interest in spin-polarized fuel has grown \cite{pace2016controlling,ciullo2016nuclear,sandorfi2017polarized,hu2023numerical} alongside progress in producing \cite{engels2016hyper,spiliotis2021ultrahigh,garcia2023,baylor2023,kannis2026production,walker2026production} and storing \cite{engels2016hyper} polarized hydrogenic species and polarization lifetime experiments \cite{didelez2011persistence,baylor2023,heidbrink2024,garcia2025,ciullo2025polarization}. Bulk nuclear polarization is expected to survive fuel injection and confinement over the required fuel residence times~\cite{Temporal_2012,heidbrink2024,baylor2023,cook2026}, with in-situ lifetime measurements planned for the DIII-D tokamak~\cite{baylor2023,garcia2023,heidbrink2024,garcia2025}. The reactivity enhancement can raise the tritium burn efficiency \cite{whyte2023,parisi2024} and so lower the required tritium throughput although polarized-fuel production and delivery at reactor-relevant throughput remains unsolved.

\begin{figure}[bp]
\includegraphics[width=0.95\columnwidth]{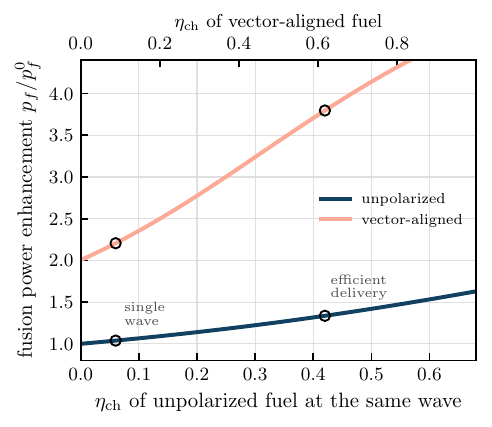}
\caption{Fusion power enhancement versus the channeling efficiency the same wave achieves on unpolarized fuel (lower x-axis) and on vector-aligned fuel (upper x-axis, $1.47$ times higher, \Cref{fig:cql}(a) for more details), from the heat transport model with $A_J=3/2$ for the aligned curve, normalized to the unpolarized, unchanneled point. Circles mark the single-wave and efficient-delivery channeling efficiencies.}
\label{fig:summary}
\end{figure}

In this work we demonstrate a constructive use of anisotropic alpha emission: enhancing energy transfer from alphas to fuel ions via alpha channeling. Most of our results come from a zero-dimensional transport model chosen for simplicity and transparency; we confirm the main results using a transport model (with radial profiles) of the published ARC V3A tokamak design equilibrium \cite{howard2026,hillesheim2026,hillesheim2026data} (\Cref{sec:arctransport}). In the vector-aligned polarization state the alphas are born anisotropic (\Cref{fig:emission}(b)), with a steady-state velocity distribution $f$ that is `population-inverted,' $\partial f/\partial v_\perp>0$, where $v_\perp$ is the velocity perpendicular to $\bm{B}$, over a broad region of velocity space, so they can drive a suitably resonant channeling wave rather than damp it, a property no isotropic slowing-down population has. The products of unpolarized thermal D-T reactions, born isotropically, cannot sustain such an inversion ($\partial f/\partial v_\perp>0$) in steady state through their velocity-space dynamics alone. For unpolarized fuel, slowing-down fills velocity space from the birth speed downward, so in steady state $f$ decreases with speed everywhere in the plasma~\cite{fisch2015}. Local inversions require orbit effects such as wide drift orbits delivering a thin ring of fast products to the tokamak edge, the source of edge ion cyclotron emission~\cite{cottrell1993,dendymcclements2015}, or the population inversion along the joint energy-position path that conventional channeling uses~\cite{fisch2015}. Polarized fuel is different: its inversion appears in the core, where the alpha power resides, for a significant fraction of the alpha birth population.

\Cref{fig:summary} previews the central result of this work. At the same channeling wave, vector-aligned (spin-polarized) fuel increases the fusion power density in three distinct ways, since the cross section factor $A_J=3/2$ makes more alphas, the temperature feedback furthers their heating, and the perpendicular birth distribution raises the channeling efficiency by about $50\%$. The vector-aligned fuel therefore roughly doubles the fusion power density before any channeling, and channeling further widens the gap, with the aligned enhancement $2.1$ times the unpolarized one at a matched single wave and $2.6$ times at matched delivery efficiency ($2.8$ at the same wave, whose aligned efficiency is $1.47$ times higher). The rest of this paper establishes each major component of the arguments to obtain the results in  \Cref{fig:summary}.

We structure this paper as follows. In \Cref{sec:control} we show how spin polarization affects the fusion reactivity and the alpha birth anisotropy. In \Cref{sec:persist} we show that the anisotropy survives the energetic phase of slowing-down. In \Cref{sec:invert} we evaluate the perpendicular free energy, the velocity-space population inversion, and the wave drive it supplies. In \Cref{sec:payoff} we calculate the resulting fusion power density increase with a transport model, and we compute the channeling efficiency. We also show the effects of partial polarization and helium-ash exhaust and its divertor pumping requirement. In \Cref{sec:hotion} we show possible further benefits of hot-ion-mode enabled by alpha channeling. In \Cref{sec:arctransport} we compute the fusion power with a profile-resolved transport model of an ARC-class plasma. We conclude in \Cref{sec:disc}. 

The appendices contain supporting calculations. \Cref{app:fp} gives the kinetic model behind the persistence and inversion results, \Cref{app:burn} the heat transport model, Appendix~\ref{app:dhe3} the extension to D-${}^{3}$He fusion, \Cref{app:arc} an ARC-like power-plant projection, \Cref{app:wave} a reduced estimate of the delivery efficiency, \Cref{app:cql} the velocity-space solver, \Cref{app:com} the orbit classification and the two-wave Monte Carlo, \Cref{app:transport} the profile-resolved transport model of the ARC plasma, and \Cref{app:cd} a simple estimate of the plasma current the channeled power could drive. Throughout, we keep the most important results in the main text and relegate the remaining details to these appendices. 

Finally, we emphasize that there are many details, some of them important, that we neglect in our modeling; this work is intended as a first step. High-fidelity modeling and experiment will be required to test the full extent of these ideas.

\section{Polarization control of the alpha birth distribution}\label{sec:control}

We begin with how the nuclear spin state controls the fusion cross section and particle emission distribution. The D-T fusion reaction produces a $\sim$14.1 MeV neutron and $\sim$3.5 MeV alpha particle via
\begin{equation}
\mathrm{d} + \mathrm{t} = \mathrm{n} + \alpha.
\end{equation}
The differential emission of alpha particles at polar angle $\theta$ from $\bm{B}$ is~\cite{kulsrud1986}
\begin{equation}
W(\theta)=1-\frac{ab}{2}+\frac{3ab}{2}\sin^2\theta+\frac{c}{4}\left(1-3\cos^2\theta\right),
\label{eq:W}
\end{equation}
where $a$ and $b$ are the deuteron and triton vector polarizations and $c$ is the deuteron tensor polarization, all dimensionless. The angle-averaged value is the cross section factor $A_J\equiv\langle W\rangle_\theta=\tfrac12\int_0^\pi W(\theta)\sin\theta\,d\theta=1+ab/2$, with $\sin\theta\,d\theta$ the solid-angle measure, which multiplies the unpolarized reaction rate, so the fusion power density of a D-T plasma is
\begin{equation}
p_f=A_J\,n_\mathrm{D}\, n_\mathrm{T}\,\langle\sigma v\rangle\,E_\mathrm{DT},
\label{eq:pf}
\end{equation}
with $n_\mathrm{D}$, $n_\mathrm{T}$ the deuteron and triton number densities, $\langle\sigma v\rangle$ the reactivity, and $E_\mathrm{DT}=17.6$ MeV the total D-T energy release per reaction. Writing the pitch variable
\begin{equation}
\xi\equiv\cos\theta=\frac{v_\parallel}{v},
\end{equation}
with $v_\parallel$ the velocity along $\bm{B}$ and $v$ the speed, the normalized birth distribution in pitch is 
\begin{equation}
f(\xi)= \frac{W(\xi)}{A_J}.
\end{equation}
Because the alphas are born at nearly a single speed, the pitch sets how their energy divides between parallel and perpendicular motion, with $\langle v_\parallel^2\rangle_\theta=\langle\xi^2\rangle_\theta\,v^2$ and $\langle v_\perp^2\rangle_\theta=\langle 1-\xi^2\rangle_\theta\,v^2$, with $\langle\cdot\rangle_\theta$ the average over the emission angle $\theta$, weighted by $f(\xi)$. We measure the anisotropy of alphas by $A_\alpha$, the ratio of mean perpendicular to parallel energy per degree of freedom,
\begin{equation}
A_\alpha\equiv\frac{\langle v_\perp^2\rangle_\theta}{2\langle v_\parallel^2\rangle_\theta}=\frac{\langle 1-\xi^2\rangle_\theta}{2\langle\xi^2\rangle_\theta},
\label{eq:aniso}
\end{equation}
which is $1$ for an isotropic birth and $2$ when the spins are vector-aligned.

\begin{figure}[bt]
\includegraphics[width=0.9\columnwidth]{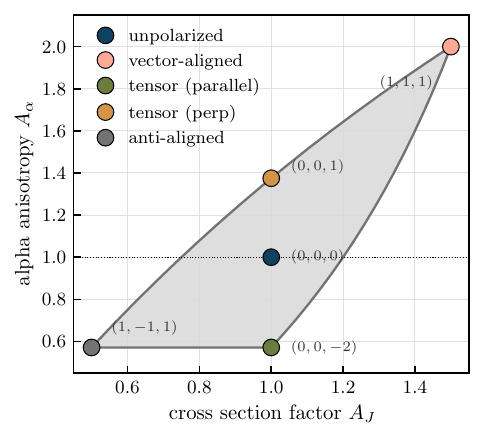}
\caption{Accessible cross section factor $A_J$ and alpha birth anisotropy $A_\alpha$ over physical deuterium and tritium polarization states (gray). Circles mark notable states with polarizations $(a,b,c)$ (\Cref{tab:states}).}
\label{fig:AJ_vs_Aalpha}
\end{figure}

The deuteron magnetic-sublevel populations constrain the vector and tensor polarizations to the region $c\in[3|a|-2,1]$, while the triton, being spin one-half, is limited to $b\in[-1,1]$. Scanning the $a$, $b$, and $c$ values gives the full range of reactivity and anisotropy in \Cref{fig:AJ_vs_Aalpha}. At maximum reactivity, $A_J=3/2$ forces $a=1\Rightarrow c=1$, giving $A_\alpha=2$. At no cost in reaction rate, $A_J=1$, the tensor term $c$ varies $A_\alpha$ from $0.57$ to $1.38$, a more perpendicular population. In this work we focus on three polarization modes: unpolarized, vector-aligned (perpendicular, $A_J=3/2$), and tensor-polarized (parallel, $A_J=1$). \Cref{tab:states} also lists the tensor-perpendicular and anti-aligned states.

\begin{table}[tb!]
\caption{Cross section factor $A_J$, alpha birth anisotropy $A_\alpha$, and perpendicular energy fraction $E_\perp\equiv\langle v_\perp^2\rangle_\theta/\langle v^2\rangle_\theta$ for notable states with polarizations $(a,b,c)$.}
\label{tab:states}
\begin{ruledtabular}
\begin{tabular}{lcccc}
state & $(a,b,c)$ & $A_J$ & $A_\alpha$ & $E_\perp$ \\
\hline
unpolarized            & $(0,0,0)$    & $1.00$ & $1.00$ & $0.667$ \\
vector-aligned         & $(1,1,1)$    & $1.50$ & $2.00$ & $0.800$ \\
tensor, perpendicular  & $(0,0,1)$    & $1.00$ & $1.38$ & $0.733$ \\
tensor, parallel       & $(0,0,-2)$   & $1.00$ & $0.57$ & $0.533$ \\
anti-aligned           & $(1,-1,1)$   & $0.50$ & $0.57$ & $0.533$ \\
\end{tabular}
\end{ruledtabular}
\end{table}

\begin{figure}[tbp]
\begin{subfigure}{0.84\columnwidth}\includegraphics[width=\linewidth]{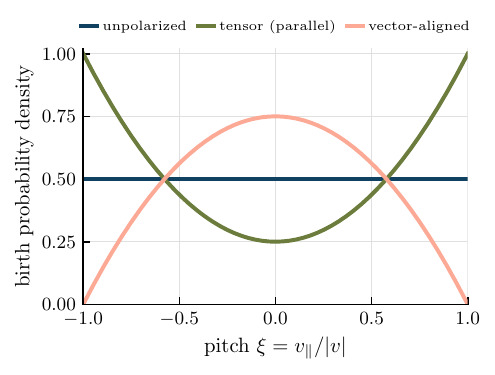}\caption{}\label{fig:birth:a}\end{subfigure}\\
\begin{subfigure}{0.84\columnwidth}\includegraphics[width=\linewidth]{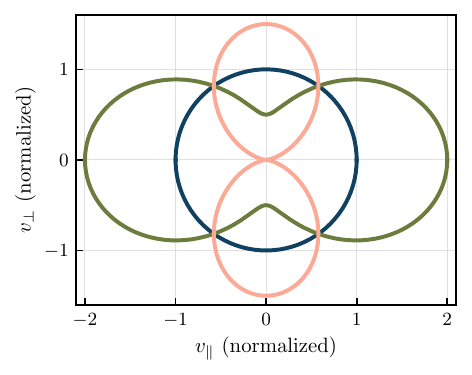}\caption{}\label{fig:birth:b}\end{subfigure}
\caption{Alpha birth distribution for the three polarization modes. (a) Pitch distribution $f(\xi)$. (b) The same as a polar plot in $(v_\parallel,v_\perp)$, the velocity components parallel and perpendicular to $\bm{B}$: the distance from the origin along a direction is the birth density of that direction, with equal particle number per mode, so the unpolarized mode is the unit circle.}
\label{fig:birth}
\end{figure}

\Cref{fig:birth} shows the alpha birth distribution for three polarization modes. In the vector-aligned mode the emission is
\begin{equation}
    W\propto\sin^2\theta,
\end{equation}
so alphas are born preferentially perpendicular to $\bm{B}$. In the tensor mode the emission is
\begin{equation}
    W\propto1+3\cos^2\theta,
\end{equation}
so they are born preferentially parallel. The unpolarized mode gives isotropic emission. For the vector-aligned mode the perpendicular energy fraction $E_\perp$ increases from $2/3$ to $4/5$ and $A_\alpha=2$. At fixed fuel density the perpendicular alpha power available to channeling therefore increases by $A_J\times(0.80/0.667)=1.8$ relative to unpolarized fuel, with a factor $1.5$ from the cross section and $1.2$ from the redistribution.

\begin{figure*}[tp]
\begin{subfigure}{0.32\textwidth}\includegraphics[width=\linewidth]{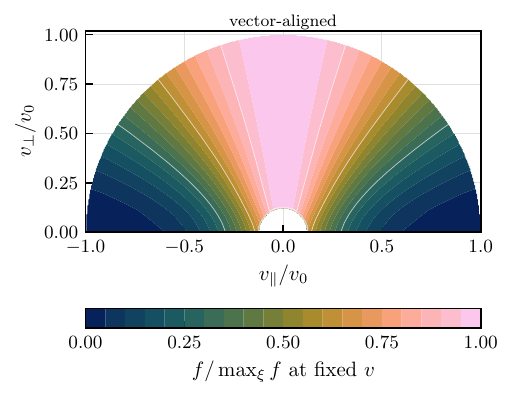}\caption{}\label{fig:fp:a}\end{subfigure}\hfill
\begin{subfigure}{0.32\textwidth}\includegraphics[width=\linewidth]{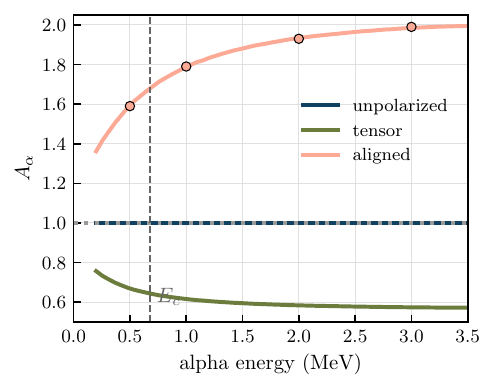}\caption{}\label{fig:fp:b}\end{subfigure}\hfill
\begin{subfigure}{0.32\textwidth}\includegraphics[width=\linewidth]{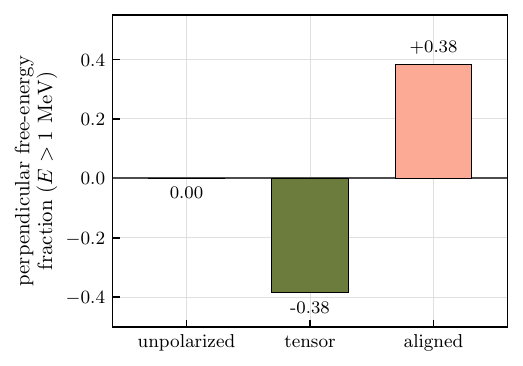}\caption{}\label{fig:fp:c}\end{subfigure}
\caption{Alpha anisotropy through slowing-down from the solver of \Cref{app:fp}. (a) Steady-state vector-aligned alpha distribution $f$ in $(v_\parallel,v_\perp)$, normalized to its pitch maximum at each speed. (b) Anisotropy $A_\alpha$ versus energy, kinetic (curves) against the analytic result of \Cref{eq:persist} (markers), with the dashed line at the critical energy $E_c$. (c) Perpendicular free-energy fraction of the $E>1$ MeV population.}
\label{fig:fp}
\end{figure*}

\section{Persistence during slowing-down}\label{sec:persist}

Channeling acts on energetic alphas, so the anisotropy must survive the alpha slowing-down in order for it to be effective. Expanding $f(\xi)$ in Legendre polynomials, the anisotropy is described by the second moment $\langle P_2\rangle_\theta$ with
\begin{equation}
P_2(\xi)=(3\xi^2-1)/2,
\end{equation}
since $\langle\xi^2\rangle_\theta=(2\langle P_2\rangle_\theta+1)/3$ turns \Cref{eq:aniso} into 
\begin{equation}
A_\alpha=\frac{1-\langle P_2\rangle_\theta}{1+2\langle P_2\rangle_\theta}.
\end{equation}
The physical effect we are studying is a competition between two collisional processes: drag lowers the alpha speed at $dv/dt=-\nu_s(v)\,v$, with $\nu_s$ the slowing-down frequency, but conserves pitch. Pitch-angle scattering randomizes the direction at the deflection frequency $\nu_d(v)$, which decays $\langle P_2\rangle_\theta$ at the rate $3\nu_d$, but conserves energy. The anisotropy survives down to a given energy if the alpha slows to it before it deflects, so what matters is the deflection accumulated per e-folding of speed, the ratio $3\nu_d/\nu_s$. Because alphas are dilute and fast, both frequencies are test-particle rates on a Maxwellian electron, deuterium, and tritium background~\cite{nrl,helandersigmar}. Light electrons drain the alpha's energy efficiently but barely turn it, since each collision transfers negligible momentum, while the fuel ions both drag and deflect. Above the critical energy $E_c$, at which the electron and ion drag rates are equal, the alpha slows mainly on the electrons and its pitch is nearly frozen. Below $E_c$ the ion collisions take over and isotropize it. Therefore, the opportunity to channel alpha energy lies at energies above $E_c$.

\begin{figure}[b]
\includegraphics[width=0.9\columnwidth]{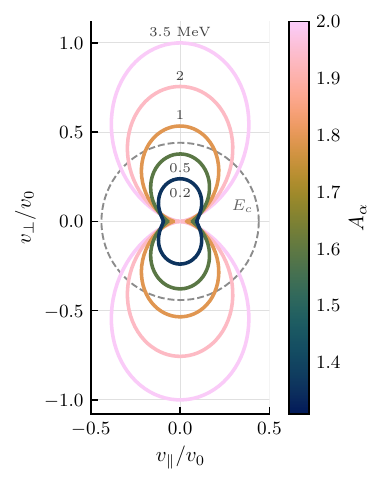}
\caption{Pitch distribution of the steady-state vector-aligned alpha population at fixed energies (labels, in MeV), from the solver of \Cref{app:fp}, as a polar plot: each curve is one constant-energy shell, and the distance from the origin at angle $\theta$ is $r(\theta)=(v/v_0)\,f(\theta)/\max_\theta f$: the pitch distribution on that shell, scaled by the shell's speed $v$ relative to the birth speed $v_0$. Color gives the shell anisotropy $A_\alpha$, and the dashed circle marks $E_c$.}
\label{fig:shapes}
\end{figure}

\begin{figure*}[tp]
\begin{subfigure}{0.32\textwidth}\includegraphics[width=\linewidth]{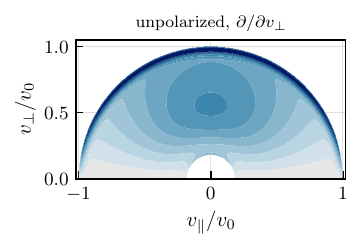}\caption{}\label{fig:inv:a}\end{subfigure}\hfill
\begin{subfigure}{0.32\textwidth}\includegraphics[width=\linewidth]{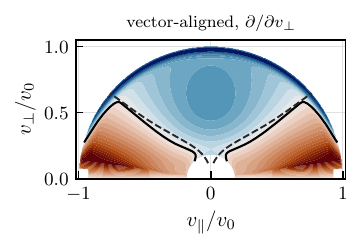}\caption{}\label{fig:inv:b}\end{subfigure}\hfill
\begin{subfigure}{0.32\textwidth}\includegraphics[width=\linewidth]{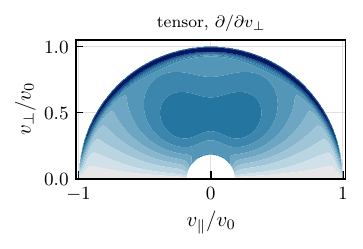}\caption{}\label{fig:inv:c}\end{subfigure}\\
\begin{subfigure}{0.32\textwidth}\includegraphics[width=\linewidth]{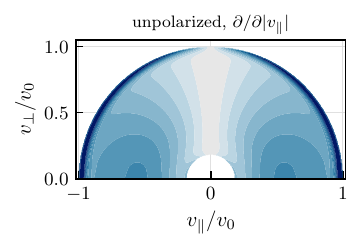}\caption{}\label{fig:inv:d}\end{subfigure}\hfill
\begin{subfigure}{0.32\textwidth}\includegraphics[width=\linewidth]{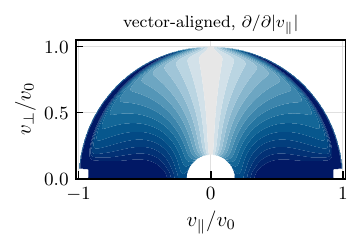}\caption{}\label{fig:inv:e}\end{subfigure}\hfill
\begin{subfigure}{0.32\textwidth}\includegraphics[width=\linewidth]{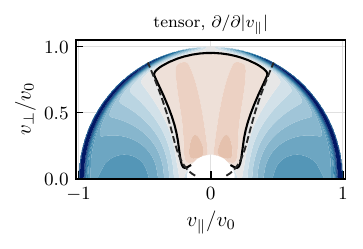}\caption{}\label{fig:inv:f}\end{subfigure}\\
\centerline{\includegraphics[width=0.52\textwidth]{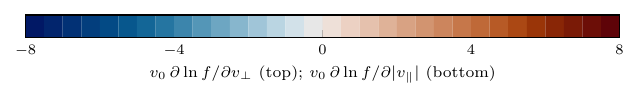}}
\caption{Velocity-space population inversion of the steady-state alpha distribution at a single spatial location, the birth source plus all slowing alphas rather than the births alone, with no spatial transport, since collisions change the alphas' velocity rather than their position, from the kinetic solver of \Cref{app:fp}: $v_0\,\partial\ln f/\partial v_\perp$ (top row) and $v_0\,\partial\ln f/\partial|v_\parallel|$ (bottom row), with $f$ the alpha distribution and $v_0$ the birth speed, for unpolarized (a,d), vector-aligned (b,e), and tensor (c,f) fuel. Red regions inside the solid zero contour are inverted. Dashed curves show the analytic boundary of \Cref{eq:invcrit}.}
\label{fig:inv}
\end{figure*}

Accumulating the $3\nu_d$ decay of $\langle P_2\rangle_\theta$ over the slowing-down history (\Cref{app:fp}) gives the surviving anisotropy fraction
\begin{equation}
\frac{\langle P_2\rangle_\theta(v)}{\langle P_2\rangle_{\theta,0}}=\exp\!\left[-\int_{v}^{v_0}\frac{3\,\nu_d(v')}{\nu_s(v')\,v'}\,dv'\right],
\label{eq:persist}
\end{equation}
where $v_0$ is the alpha birth speed and $\langle P_2\rangle_{\theta,0}$ the birth value. For a background with equal ion and electron temperatures of $15$ keV, $E_c\approx0.68$ MeV in the drag-rate convention (the burn model's convention gives a slightly smaller value, \Cref{app:burn}). The alpha is born with birth energy $\mathcal{E}_0=3.5$ MeV, far above $E_c$, so the alpha loses the first $\sim80\%$ of its energy in the electron-drag regime with its pitch nearly frozen. This is a large energy range where channeling can act on a still-anisotropic population, and it covers almost the full alpha energy (\Cref{fig:fp}(b), from the solver of \Cref{app:fp}). The vector-aligned mode has $A_\alpha=1.99$, $1.93$, and $1.79$ at $3$, $2$, and $1$ MeV, isotropizing quickly only below $E_c$. \Cref{fig:shapes} draws these shapes directly, each at its own speed and colored by its shell $A_\alpha$, which follows the decay of \Cref{eq:persist}: the birth anisotropy survives nearly unchanged down to $1$ MeV. The steady-state slowing-down population averages $A_\alpha=1.92$ above $1$ MeV and $1.80$ above $0.2$ MeV, so the anisotropy persists across the energy range that channeling exploits. The strong scattering below $E_c$ therefore reaches each alpha only after it has given up most of its energy.

\begin{figure*}[tp]
\begin{subfigure}{0.24\textwidth}\includegraphics[width=\linewidth]{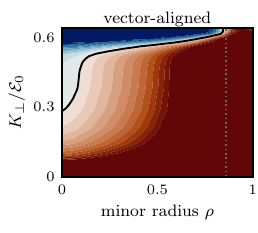}\caption{}\label{fig:chancond:a}\end{subfigure}\hfill
\begin{subfigure}{0.24\textwidth}\includegraphics[width=\linewidth]{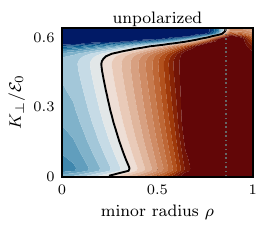}\caption{}\label{fig:chancond:b}\end{subfigure}\hfill
\begin{subfigure}{0.24\textwidth}\includegraphics[width=\linewidth]{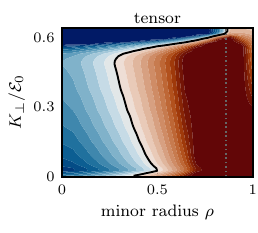}\caption{}\label{fig:chancond:c}\end{subfigure}\hfill
\begin{subfigure}{0.24\textwidth}\includegraphics[width=\linewidth]{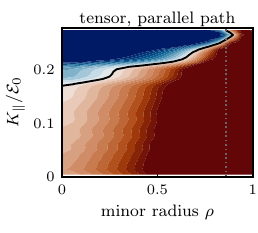}\caption{}\label{fig:chancond:d}\end{subfigure}\\
\centerline{\includegraphics[width=0.36\textwidth]{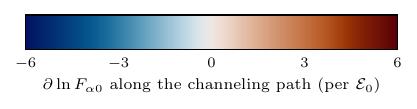}}
\caption{The amplification condition of \Cref{eq:chancond}~\cite{ochsfisch2021} evaluated on the no-wave steady state of \Cref{app:cql}: the logarithmic derivative of the alpha distribution along the channeling path (energy derivative plus spatial derivative at the path slope), with an assumed fusion-source profile $\propto(1-\rho^2)^4$ in normalized minor radius $\rho$. Panels (a)-(c) take the perpendicular path at the parallel resonance $v_\parallel=0.6\,v_0$, with $v_0$ the alpha birth speed and $K_\perp$ the perpendicular energy, for (a) vector-aligned, (b) unpolarized, and (c) tensor fuel. Panel (d) takes the parallel, Landau-resonant path at fixed $v_\perp=0.85\,v_0$ for tensor fuel, with $K_\parallel$ the parallel energy. Red regions amplify the wave and the solid curve marks zero. The dotted line marks where the solver removes alphas to represent loss at the plasma edge, so the gradients beyond it come from that removal rather than from the drive.}
\label{fig:chancond}
\end{figure*}

\section{Velocity-space free energy and population inversion}\label{sec:invert}
\subsection{Anisotropy as free energy}
Having shown that the anisotropy survives, we now show how anisotropic alphas interact with waves. Alpha channeling extracts the alphas' high birth energy whatever their pitch distribution, but an anisotropic population offers, in addition, free energy that can amplify the wave. Setting aside the spatial-gradient route of standard channeling, a perpendicular-resonant wave gains net energy from the pitch structure only when the perpendicular motion holds more energy than an isotropic distribution would. That excess is $\tfrac12\langle v_\perp^2\rangle_\theta-\langle v_\parallel^2\rangle_\theta$ per unit mass, the combination of perpendicular and parallel energies that vanishes for an isotropic distribution where $\langle v_\perp^2\rangle_\theta=2\langle v_\parallel^2\rangle_\theta$. Free energy in general is whatever part of a distribution's kinetic energy some rearrangement can pass to a wave; the alphas' high birth energy itself qualifies, and standard channeling extracts it through the radial gradient~\cite{fisch1992}. Here we use free energy in the restricted sense of the anisotropy excess, the energy available beyond what the same alphas would offer if isotropic. Collisional relaxation of alphas to isotropy releases none of this excess to the plasma; only a wave whose resonance selects the perpendicular motion can extract it. For an isotropic birth it vanishes and stays zero, since drag acts on the speed and pitch-angle scattering only relaxes toward isotropy. For the vector-aligned birth it is $0.20\,v^2$, a fraction $0.4$ of the kinetic energy, decreasing to $0.17\,v^2$ over the slowing-down population above $0.2$ MeV. For the tensor mode it is negative, so a perpendicular wave is damped rather than driven. Direct simulation of the polarized-D-T magnetoacoustic instability by Cook et al. is consistent with this picture, with vector alignment raising the wave growth rate by about $20\%$ and the tensor mode lowering it to two-thirds~\cite{cook2026}. The growth of related ring-driven modes is likewise set mainly by the perpendicular velocity range~\cite{kong2023}.

A quasilinear Fokker-Planck calculation (\Cref{app:fp})~\cite{cianfrani2018} confirms this. The steady-state anisotropy reproduces the analytic slowing-down result, and the perpendicular free-energy fraction of the energetic population is $+0.38$ for vector-aligned, $0.00$ for unpolarized, and $-0.38$ for tensor (\Cref{fig:fp}). A broadband perpendicular wave added to the same kinetic equation net-damps for every mode (\Cref{app:fp}), so extraction of the alpha energy requires the resonant, energy-space diffusion path of alpha channeling~\cite{fisch2015}, with the polarization-set anisotropy amplifying that wave. The channeling wave does not necessarily need to be self-excited, since in conventional alpha channeling it is launched externally, but the free energy could reduce the power needed to sustain it.

\subsection{Population inversion}

We next locate where in velocity space a wave can reach this free energy. Population inversion is the feature of the alpha distribution function that allows alphas to drive a resonant wave rather than damp it, since a wave grows or damps according to the sign of the distribution gradient along its diffusion path. Ochs and Fisch give the wave amplification condition~\cite{ochsfisch2021},
\begin{equation}
\left(\frac{\partial}{\partial K}+\frac{\bm{k}\times\hat{\bm{b}}}{m_\alpha\,\omega\,\Omega}\cdot\frac{\partial}{\partial\bm{X}}\right)F_{\alpha0}>0,
\label{eq:chancond}
\end{equation}
with $F_{\alpha0}$ the alpha gyrocenter distribution, $K$ the perpendicular kinetic energy, $\bm{X}$ the gyrocenter position, $m_\alpha$ the alpha mass, $\Omega$ the alpha cyclotron frequency, $\hat{\bm{b}}$ the field direction, and $\bm{k}$ the wavenumber.

Conventional channeling satisfies \Cref{eq:chancond} through the second, spatial term (the populated core against the empty edge), while the local energy derivative is negative or zero. For a magnetized cyclotron-resonant wave the condition applies at the parallel resonance, since integrating over $v_\parallel$ would average the pitch structure away. A population inversion makes the first term positive, since $\partial F_{\alpha0}/\partial K>0$ is the energy form of $\partial f/\partial v_\perp>0$. Therefore vector-aligned fuel can improve on both terms in \Cref{eq:chancond}, with the fuel polarization determining whether the first term contributes. The inversion is visible in the velocity-space gradients that control wave growth (\Cref{fig:inv}). For unpolarized fuel both $\partial f/\partial v_\perp$ and $\partial f/\partial|v_\parallel|$ are negative, since slowing-down fills velocity space from the birth speed downward, and every resonant wave damps. The anisotropic birth overturns this wherever the pitch dependence of $f$ steepens faster than the speed dependence decays. The vector-aligned distribution rises with $v_\perp$ wherever the pitch angle satisfies
\begin{equation}
\frac{2\xi^2}{1-\xi^2} > \frac{3v^3}{v^3+v_c^3},
\end{equation}
with $v_c$ the speed at $E_c$ (derived in \Cref{app:fp}), an outer-pitch region $|\xi|\gtrsim0.76$ near the birth speed, and the solver of \Cref{app:fp} confirms $\partial f/\partial v_\perp>0$ over $46\%$ of the occupied $(v_\parallel,v_\perp)$ space, tracking this boundary (\Cref{fig:inv}(b)). 

The tensor mode is also population-inverted, $\partial f/\partial|v_\parallel|>0$ for $|\xi|\lesssim0.48$, over $35\%$ of that space (\Cref{fig:inv}(f)). Both are area measures that weight sparsely and densely populated regions equally. Each mode is inverted along the direction of its excess birth energy, and only there. Fusion products of unpolarized thermal fuel, which are born isotropically, are known to invert only through orbit effects, most clearly where centrally born alphas on wide trapped orbits reach the outer midplane edge and form the locally inverted ring distributions that drive ion cyclotron emission, an inversion confined to a thin edge layer and a small subpopulation~\cite{cottrell1993,dendymcclements2015,cookdendy2017}. The polarization-set inversion is instead a property of the whole birth population, in the core where the alpha power resides.

\Cref{fig:chancond} shows the left side of \Cref{eq:chancond} on the no-wave steady state (\Cref{app:cql}), at the parallel resonance $v_\parallel=0.6\,v_0$ and along the channeling path. On axis ($\rho = 0$) the spatial term vanishes at the peak of the birth profile, so the sign there is set by the energy term alone. For unpolarized and tensor fuel the on-axis column is nowhere positive, and the inner third of the plasma damps the wave. Vector alignment turns $58\%$ of the on-axis column positive and pushes the zero contour to the axis, so the wave gains where the alpha power resides. On the parallel, Landau-resonant path of its own scheme (\Cref{fig:chancond}(d)), the tensor mode instead turns $62\%$ of the column positive while the other modes are nowhere positive. Each polarized mode drives the wave class matched to its inversion, and unpolarized fuel drives neither in the core.

\subsection{Wave drive}
A wave gains energy from the inversion only if its resonant particles are the inverted ones. Evaluating the Kennel-Engelmann drive on the steady state, the drive at the first cyclotron harmonic is positive for vector-aligned fuel over roughly half of the wave-parameter range, at parallel resonances $v_{\parallel,\rm res}\gtrsim0.2\,v_0$, and negative at every point for unpolarized and tensor fuel, so the polarization state alone decides whether an amplification region exists (\Cref{app:fp}). Against bulk damping the aligned-alpha drive reaches $14\%$ of the electron Landau damping at reactor alpha densities. We therefore read this drive as lowering the power needed to sustain a wave tuned into it, with the bulk of the channeling proceeding by the resonant radial path of \Cref{app:cql}; the drive is calculated from the alpha distribution before any wave is applied, and whether the reduction is realized depends on the specific wave's mode structure and damping (\Cref{app:fp}).

\section{Fusion power enhancement}\label{sec:payoff}
\subsection{Heat transport model}
We have shown that the alpha anisotropy persists under collisions (\Cref{sec:persist}) and can feed a wave (\Cref{sec:invert}), so we now calculate the effect on the fusion power. Spin alignment raises the fusion power density $p_f$ (\Cref{eq:pf}) directly through $A_J$, and indirectly by channeling alpha power to the ions to raise $T_i$ and hence $\langle\sigma v\rangle$. We quantify both effects with a zero-dimensional heat transport model where channeling enters as a rerouting of the alpha power. We mainly use the zero-dimensional model throughout this paper. In \Cref{sec:arctransport} we test its conclusions using a transport model with radial profiles of the ARC V3A design equilibrium.

\begin{figure}[tbp]
\begin{subfigure}{\columnwidth}\includegraphics[width=0.95\linewidth]{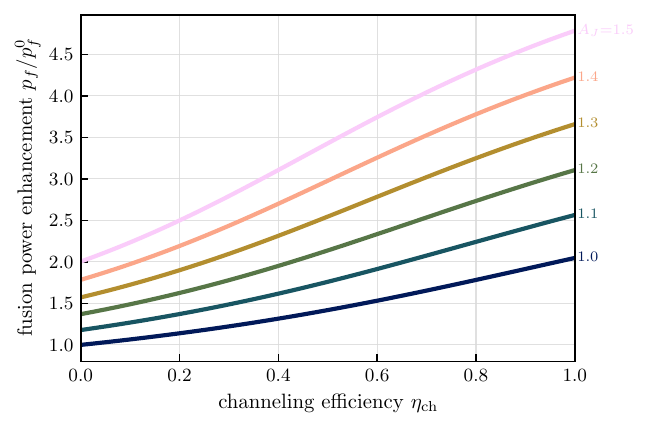}\caption{}\label{fig:gain:a}\end{subfigure}\\
\begin{subfigure}{\columnwidth}\includegraphics[width=0.95\linewidth]{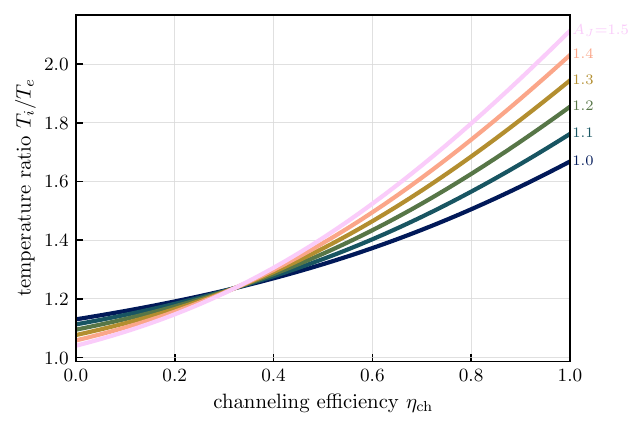}\caption{}\label{fig:gain:b}\end{subfigure}
\caption{(a) Fusion power density enhancement versus the channeling efficiency $\eta_{\rm ch}$ for the cross section factor $A_J$ from $1$ to $3/2$, from the heat transport model.  (b) The ion-to-electron temperature ratio $T_i/T_e$ for the same scan.}
\label{fig:gain}
\end{figure}

\begin{figure*}[tp]
\begin{subfigure}{0.48\textwidth}\includegraphics[width=\linewidth]{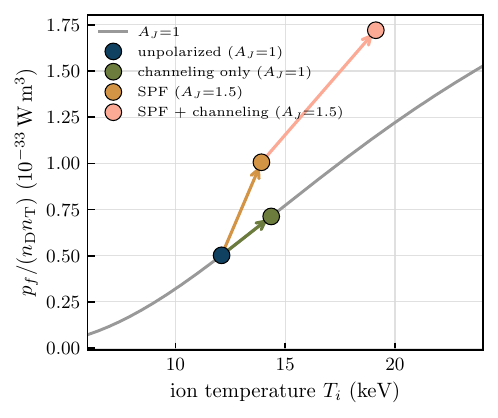}\caption{}\label{fig:nonlin:a}\end{subfigure}\hfill
\begin{subfigure}{0.48\textwidth}\includegraphics[width=\linewidth]{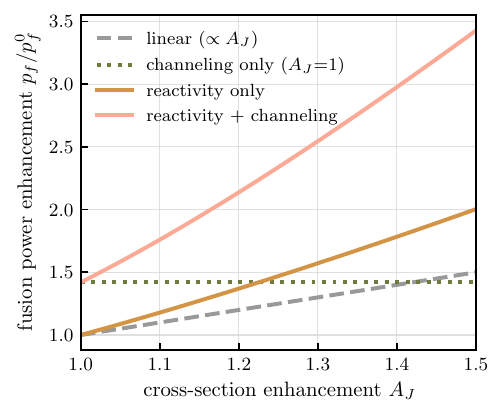}\caption{}\label{fig:nonlin:b}\end{subfigure}
\caption{(a) Operating points on $p_f/(n_\mathrm{D} n_\mathrm{T})=A_J\langle\sigma v\rangle(T_i)E_\mathrm{DT}$ (\Cref{eq:pf}, with $p_f$ the fusion power density, $n_\mathrm{D}$, $n_\mathrm{T}$ the fuel densities, $A_J$ the cross section factor, $\langle\sigma v\rangle$ the reactivity, and $E_\mathrm{DT}$ the reaction energy), where each effect raises the ion temperature $T_i$ along the $A_J{=}1$ curve and $A_J=1.5$ raises the spin-polarized-fuel (SPF) cases above it. (b) Gain versus $A_J$ from the heat transport model, with channeling alone at channeling efficiency $\eta_{\rm ch}=0.5$ ($A_J=1$) dotted.}
\label{fig:nonlin}
\end{figure*}

At a given ion temperature $T_i$ and electron temperature $T_e$, only a fraction $f_i$ of the alpha power $P_\alpha$ reaches the ions and the remainder $(1-f_i)$ heats the electrons, because the alphas are born well above the critical energy and first drag on electrons. At the operating points of the burn below, $f_i\approx0.2$. A channeling wave extracts a fraction $\eta_{\rm ch}$ of the alpha power before it slows collisionally and delivers it to the ions, while the remainder splits collisionally, so the ion-heating power from alphas is
\begin{equation}
P_{\alpha\to i}=P_\alpha\big[f_i+\eta_{\rm ch}\,(1-f_i)\big],\qquad \eta_{\rm ch}=\varepsilon_w E_\perp,
\label{eq:etach}
\end{equation}
and the electrons receive the rest (\Cref{app:burn}). The channeling fraction $\eta_{\rm ch}$ is the share of the total alpha power delivered to the ions, since \Cref{eq:etach} rearranges to $P_{\alpha\to i}=P_\alpha[\eta_{\rm ch}+(1-\eta_{\rm ch})f_i]$, and for the perpendicular-resonant single-wave we focus on it factors into the channelable perpendicular fraction $E_\perp$ and a delivery efficiency $\varepsilon_w$ (\Cref{app:wave}). 

The factor $E_\perp$ is a bound only in a uniform field. There, a perpendicular-resonant wave changes just the perpendicular velocity, so once it has drained an alpha's perpendicular energy the parallel energy sits beyond its reach, and the extraction per alpha stops at the birth fraction $E_\perp$. A trapped alpha in a torus escapes this bound, with its energy moving back and forth between parallel and perpendicular motion along the trapped orbit, and at the turning points where the parallel velocity vanishes, all of its energy is perpendicular; a wave resonant at these turning points finds perpendicular energy on every bounce and can extract nearly the full birth energy (\Cref{app:com}). Vector alignment helps twice here, raising $E_\perp$ itself and raising the trapped birth fraction, from $30\%$ to $43\%$ at the reference reactor of \Cref{app:com}. The absolute trapped fractions scale with the aspect ratio, but the aligned-to-unpolarized ratio stays near $1.4$ from conventional to spherical tokamaks, since the birth pitch weighting sets it rather than the geometry. We take $\eta_{\rm ch}=0.5$ for the vector-aligned mode ($\varepsilon_w=0.625$, $E_\perp=0.8$) and $0.3$ for the tensor mode, whose less-developed parallel-acting scheme channels the parallel fraction $E_\parallel=0.47$ instead and rounds to the same value. Both are heuristic rather than rigorously derived, though they fall in the range reported for channeling simulations in prior work, which have been shown to extract more than $60\%$ of the resonant alpha energy~\cite{zhmoginov2008,ochs2022}. 

Reaching these high channeling efficiencies requires a multi-harmonic or broadband scheme, and we assume alignment's larger efficiency persists from the single-wave to the multi-harmonic or broadband schemes. The referenced simulations also count energy per resonant alpha, so using their numbers for a fraction of the total alpha power assumes the wave spectrum reaches most of the population.
\begin{figure*}[tp]
\begin{subfigure}{0.46\textwidth}\includegraphics[width=\linewidth]{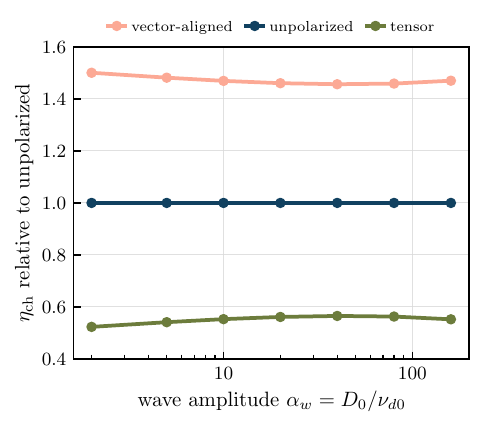}\caption{}\label{fig:cql:a}\end{subfigure}\hfill
\begin{subfigure}{0.46\textwidth}\includegraphics[width=\linewidth]{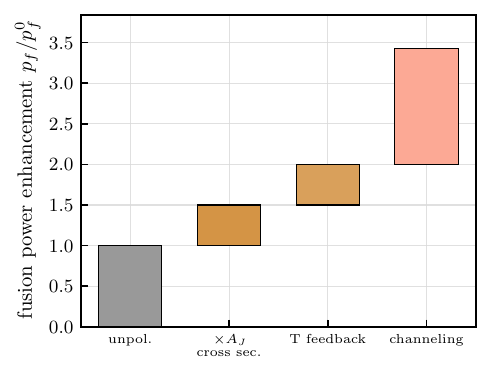}\caption{}\label{fig:cql:b}\end{subfigure}
\caption{First-principles channeling from the velocity-space solver (\Cref{app:cql}). (a) Channeling efficiency relative to unpolarized fuel versus wave amplitude $\alpha_w=D_0/\nu_{d0}$, with $D_0$ the quasilinear diffusion rate at the birth speed and $\nu_{d0}$ the deflection frequency there. (b) The vector-aligned enhancement broken into its contributions, from the heat transport model at the adopted $\eta_{\rm ch}=0.5$.}
\label{fig:cql}
\end{figure*}

A reduced solver of the resonant operator (\Cref{app:wave}) puts $\varepsilon_w$ between about $0.1$ and $0.5$ at moderate drive, reaching our adopted $\varepsilon_w=0.625$ only when the wave drive exceeds the collision rate by a factor of ten. The perpendicular free energy, a fraction $0.38$ of the aligned alphas' energy (\Cref{eq:ufree}), feeds the wave and lowers the antenna power~\cite{cookdendy2017}.

\begin{figure}[tbp]
\includegraphics[width=0.95\columnwidth]{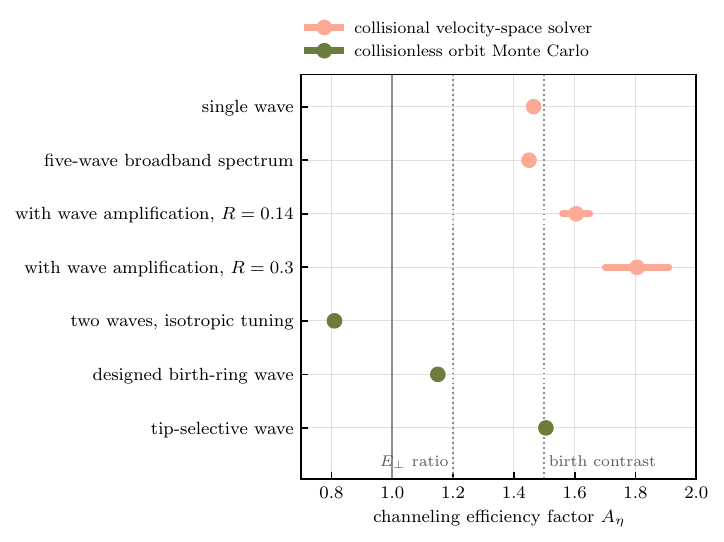}
\caption{The channeling efficiency factor $A_\eta=\eta_{\rm ch}^\mathrm{aligned}/\eta_{\rm ch}^\mathrm{isotropic}$ for the calculations of this work. Upper rows: the collisional velocity-space solver of \Cref{app:cql} with a single wave, with five broadband waves at staggered resonant velocities, and at the amplitude sustained when the anisotropy offsets a fraction $R$ of the wave damping (\Cref{fig:drive}(b)). Lower rows: the collisionless orbit Monte Carlo of \Cref{app:com} with the two-wave scheme at its isotropic-fuel tuning, with the single wave designed for aligned births, and with that wave restricted to deeply trapped orbits.}
\label{fig:multiplier}
\end{figure}
Our heat transport model (see \Cref{app:burn} for more details) evolves $T_i$ and $T_e$ with the split alpha heating, channeling, ion-electron exchange, radiation, auxiliary heating, and power-degraded confinement, so extra alpha heating raises $T_i$ and $\langle\sigma v\rangle$ self-consistently~\cite{parisi2024} (\Cref{fig:nonlin}(a)). We omit more complex, but ultimately necessary, effects such as profiles, geometry, and impurities. We set its parameters so that the $50\%$ cross-section enhancement~\cite{kulsrud1982,kulsrud1986}, compounded by this feedback, roughly doubles the fusion power, a choice that sets the operating point rather than predicting the doubling. 

Channeling can also drive the hot-ion state that increases fusion output at fixed pressure~\cite{fisch1992} (\Cref{fig:gain}(b)). The reactivity boost and feedback give a factor $2.0$ (\Cref{tab:ign}, \Cref{fig:gain}), consistent with the increases projected by Smith et al.~\cite{smith2018}. Channeling raises this to $2.2$ at the single-wave $\eta_{\rm ch}\approx0.09$ (\Cref{app:cql}), $3.4$ at $\eta_{\rm ch}=0.5$ (\Cref{fig:nonlin}(b)), and about $4.0$ at $\eta_{\rm ch}\approx0.7$~\cite{zhmoginov2008}. 

The single-wave efficiency of $\eta_{\rm ch}=0.09$ that we compute for vector-aligned fuel is a lower bound on $\eta_{\rm ch}$: two-wave and mirror calculations on isotropic fuel reach $\eta_{\rm ch}\approx0.5$ to $0.7$~\cite{herrmann1997,fischherrmann1994,zhmoginov2008}, and whatever efficiency a scheme achieves on isotropic fuel, the perpendicular-born population raises it by the channeling efficiency factor
\begin{equation}
A_\eta\equiv\eta_{\rm ch}^\mathrm{aligned}/\eta_{\rm ch}^\mathrm{isotropic},
\end{equation}
with $A_\eta \approx1.5$ at the same wave (\Cref{fig:cql}(a)). This increase persists with a broadband spectrum: we find five waves at staggered resonant velocities in the velocity-space solver preserve $A_\eta$ at $1.4$ to $1.5$ while raising the absolute efficiency (\Cref{app:cql}). The factor $A_\eta$, not any single efficiency value, is the quantity we establish in this paper. We plot $A_\eta$ across our calculations in \Cref{fig:multiplier}. The velocity-space solver finds $A_\eta$ stays close to $1.5$ from a single wave to five, and wave amplification by the anisotropy free energy raises it to $1.6$ to $1.9$ as the alpha drive offsets a growing fraction $R$ of the wave damping (\Cref{fig:drive}(b)). The collisionless orbit calculations give $A_\eta=0.8$ for the two-wave scheme (at its isotropic tuning), $1.15$ for the wave designed for coverage, and $1.5$ for a tip-selective wave that interacts only with deeply trapped alphas. Restricting the wave to that population lowers the total diverted power, but it raises $A_\eta$ to the largest value any collisionless scheme can reach: the factor $3/2$ by which aligned births outnumber unpolarized ones at perpendicular pitch.

The fusion-power enhancements above hold the density fixed and let the pressure rise with the extra heating. If we instead hold the total heating power, and with it the stored energy, fixed by lowering the auxiliary power, the fusion power reaches $1.3$ times the unpolarized, unchanneled value without channeling and $1.6$ times at $\eta_{\rm ch}=0.5$, with the auxiliary power falling from $0.5$ to $0.32$ and $0.18$ MW\,m$^{-3}$: at fixed pressure, part of the benefit converts into reduced drive. Channeling unpolarized fuel at the same delivery efficiency raises its fusion power $1.3$ times, so dividing the aligned enhancement by the unpolarized one gives the polarization-attributable factor, $2.6$ at matched channeling efficiency and $2.1$ at the matched single-wave values. The doubling requires no channeling at all, and channeling builds on it in proportion to the achievable efficiency. In \Cref{app:burn} we vary the model parameters and find $10$ to $90\%$ ranges of $1.6$ to $2.9$ for the unchanneled doubling and $2.1$ to $7.5$ for the channeled enhancement for these two enhancements (\Cref{fig:sens}).

\begin{table}[b]
\caption{Fusion power density enhancement relative to the unpolarized plasma, from the heat transport model, as the channeling efficiency $\eta_{\rm ch}$ improves, with $A_J$ the cross section factor.}
\label{tab:ign}
\begin{ruledtabular}
\begin{tabular}{lccc}
case & $A_J$ & $\eta_{\rm ch}$ & power enhancement \\
\hline
unpolarized                          & $1.0$ & $0.0$ & $1.0$ \\
unpolarized $+$ efficient channeling & $1.0$ & $0.42$ & $1.3$ \\
reactivity only                      & $1.5$ & $0.0$ & $2.0$ \\
$+$ single-wave channeling           & $1.5$ & $0.09$ & $2.2$ \\
$+$ efficient channeling             & $1.5$ & $0.5$ & $3.4$ \\
$+$ literature-scale channeling      & $1.5$ & $0.7$ & $4.0$ \\
\end{tabular}
\end{ruledtabular}
\end{table}

\subsection{Channeling efficiency from velocity space}

The $\eta_{\rm ch}$ values above were assumed rather than derived, so we now compute the channeling efficiency directly from velocity space. 

We solve the alpha distribution in speed, pitch, and minor radius with a quasilinear Fokker-Planck model (see \Cref{app:cql}), which combines the linearized collision operator, the resonant Kennel-Engelmann operator for a cyclotron-harmonic wave~\cite{kennelengelmann1966}, and the radial transport the wave imposes through $P_\phi$~\cite{fisch1992,fisch2015}. In quasilinear theory the resonant alphas take many small, uncorrelated kicks from the wave, so the wave enters the kinetic equation as a velocity-space diffusion operator whose strength is proportional to the wave intensity~\cite{kennelengelmann1966,cql3d}, an approximation valid when each kick only weakly perturbs the orbit. We do not assume the extraction is perpendicular: the finite-Larmor-radius coupling makes it so, and more than $99\%$ of the extracted energy is perpendicular. The calculation confirms that $\eta_{\rm ch}$ is largest for vector-aligned fuel and smallest for the parallel tensor mode (\Cref{fig:cql}(a)), and finds the anisotropy enhancement, in which the birth anisotropy raises the per-alpha channeling efficiency by $A_\eta=1.47$ ($1.20$ from the larger perpendicular fraction $E_\perp$, $1.23$ from better resonance overlap), while the parallel tensor mode channels at only $0.55$ times the unpolarized rate. \Cref{fig:cql}(b) breaks the vector-aligned enhancement into its contributions, this channeling factor together with the cross-section boost and the temperature feedback those alphas drive. 

The single-wave efficiency is modest, about $0.09$, limited by the coupling weakening as the alpha is channeled to low $v_\perp$, so the larger $\eta_{\rm ch}$ used above needs waves that cover more of velocity space. We tested multi-wave spectra directly (\Cref{app:cql}). A second perpendicular harmonic at higher $k_\perp$ raises $\eta_{\rm ch}$ at fixed per-wave amplitude, but it saturates at the same value. Waves that reach down to the slowed population heat it instead. The established route to larger $\eta_{\rm ch}$ is therefore likely a second, $\mu$-conserving wave~\cite{herrmann1997}. The factor $A_\eta$ varies by less than $3\%$ across the factor-of-$80$ wave amplitude range of \Cref{fig:cql}(a). In position space, alphas born near the magnetic axis arrive at larger radius progressively colder, because the wave's quasilinear path ties each outward step to the perpendicular energy it extracts (see \Cref{app:cql}). The polarization also relocates the births in orbit space, loading the region reached by each wave of the two-wave scheme of Herrmann and Fisch~\cite{herrmann1997}, and at wave amplitudes tuned for isotropic fuel the diverted power falls as the polarization rises (see \Cref{app:com}). Just retuning the amplitudes does not recover it because the interaction probability is already saturated for the high-$\mu$ aligned births. However, redesigning the wave does: in the collisionless Monte Carlo of \Cref{app:com} we design a single wave whose quasilinear path follows the aligned births, which cluster near a single point of the constants-of-motion space because a perpendicular birth holds its full energy in magnetic moment, so it diverts $60\%$ of the alpha power, more than the two-wave scheme achieves on isotropic fuel.

These results complete \Cref{fig:summary} from the introduction: with the $1.47$ efficiency factor we calculated here, the enhancement of aligned fuel is $2.1$ times the unpolarized one using a single-wave scheme and $2.6$ times at matched delivery efficiency ($2.8$ at the same wave, \Cref{fig:summary}).

\subsection{Partial polarization and the accessible states}

\begin{figure}[tbp]
\includegraphics[width=0.95\columnwidth]{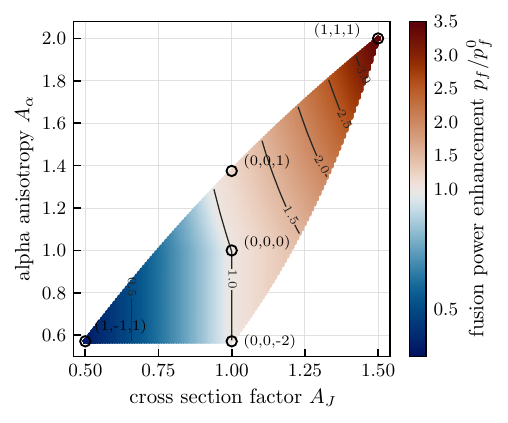}
\caption{Fusion power enhancement over the accessible polarization states, from the heat transport model along the polarization-scaled channeling path $\eta_{\rm ch}=\varepsilon_w E_\perp$ (see text), with $\eta_{\rm ch}$ the channeling efficiency, $\varepsilon_w$ the delivery efficiency, and $E_\perp$ the perpendicular energy fraction. The color scale diverges about unity enhancement. Circles mark the notable states of \Cref{tab:states} with polarizations $(a,b,c)$.}
\label{fig:payoff}
\end{figure}

The enhancements so far assume fully polarized fuel. Realistic fuel may be partially polarized, so we scale $a=b=c=p$ from $0$ to $1$, giving 
\begin{equation}
A_J=1+p^2/2,
\end{equation}
an anisotropy falling from $A_\alpha=2$ to $1$, and a polarization-scaled $\eta_{\rm ch}=\varepsilon_w E_\perp$, an assumption that scales $\varepsilon_w$ with the alphas' free energy so channeling is counted only where the anisotropy could feed the wave, vanishing at $p=0$. An externally driven wave at fixed $\varepsilon_w$ would keep a finite $\eta_{\rm ch}$ even at $p=0$. The burn then gives about $3.4$ times unpolarized at $p=1$ and $2.1$ at $p=0.75$ ($A_J=1.28$, $A_\alpha=1.62$), the $a{=}b{=}c$ diagonal of \Cref{fig:payoff}. The benefit persists and rises with achievable polarization. 

We repeat this burn calculation across all accessible polarization states. In \Cref{fig:payoff} we color every reachable $(A_J,A_\alpha)$ state by its burn enhancement along the same polarization-scaled path. The enhancement ranges from $0.33$ for anti-aligned fuel, through $1$ for unpolarized fuel, to $3.4$ at full vector alignment. The deuteron tensor term alone, at no cost in reaction rate, ranges from $1.0$ to $1.16$, the perpendicular tensor state gaining $16\%$ from anisotropy-scaled channeling under this assumption.

\subsection{Ash exhaust}

Channeling also reduces core helium because the wave ejects the alphas before they can thermalize. We extend the heat transport model with a heuristic balance for the core helium (\Cref{app:burn}). This section draws heavily from notation in~\cite{whyte2023} and~\cite{parisi2024}.

We split the core helium density into fast alphas of density $n_\mathrm{fast}$ and thermalized ash of density $n_\mathrm{ash}$, with total 
\begin{equation}
n_\mathrm{He}=n_\mathrm{fast}+n_\mathrm{ash}.
\end{equation}
Every birth enters the fast population. The wave moves the fraction $F_\mathrm{ej}$ to the edge during extraction, quickly compared with slowing down, so it doesn't accumulate in the core, and the remainder thermalizes into the ash,
\begin{equation}
\dot n_\mathrm{fast}=\dot n_\alpha(1-F_\mathrm{ej})-\frac{n_\mathrm{fast}}{\tau_\mathrm{sd}},\qquad
\dot n_\mathrm{ash}=\frac{n_\mathrm{fast}}{\tau_\mathrm{sd}}-\frac{n_\mathrm{ash}}{\tau_\mathrm{ash}},
\label{eq:nHethermal}
\end{equation}
where the alpha birth rate is
\begin{equation}
\dot n_\alpha = \frac{p_f}{E_\mathrm{DT}},
\label{eq:ndotalpha}
\end{equation}
$\tau_\mathrm{sd}$ is the alpha slowing-down time, $\tau_\mathrm{ash}=2.5\,\tau_E$ the core residence time of a thermalized helium ion, and $\tau_E$ the energy confinement time. We chose $\tau_\mathrm{ash}$ so the standard operating point has $n_\mathrm{ash}/n_e\approx2\%$, total helium $n_\mathrm{He}/n_e\approx2.6\%$. Helium dilutes the fuel through quasineutrality at fixed $n_e$,
\begin{equation}
n_e = n_\mathrm{D} + n_\mathrm{T} + 2 n_\mathrm{He}.   
\end{equation}

In steady state \Cref{eq:nHethermal} gives
\begin{equation}
n_\mathrm{fast}=\dot n_\alpha(1-F_\mathrm{ej})\,\tau_\mathrm{sd}.
\end{equation}
With $\tau_\mathrm{sd}\approx0.3$ to $0.4$ s against $\tau_\mathrm{ash}=1.2$ to $2.5$ s across our operating points, $n_\mathrm{fast}/n_\mathrm{ash}=\tau_\mathrm{sd}/\tau_\mathrm{ash}$ is an eighth to a third and $n_\mathrm{fast}$ stays below half a percent of $n_e$. We include both populations in the dilution. The ash balance then gives
\begin{equation}
n_\mathrm{ash} = \tau_\mathrm{ash} \dot n_\alpha(1-F_\mathrm{ej}).
\label{eq:nashss}
\end{equation}

\begin{figure}[tbp]
\begin{subfigure}{\columnwidth}\includegraphics[width=0.95\linewidth]{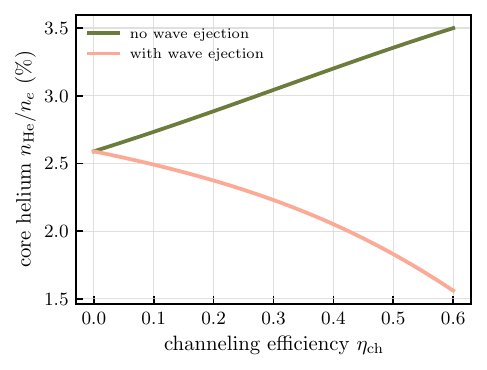}\caption{}\label{fig:ash:a}\end{subfigure}\\
\begin{subfigure}{\columnwidth}\includegraphics[width=0.95\linewidth]{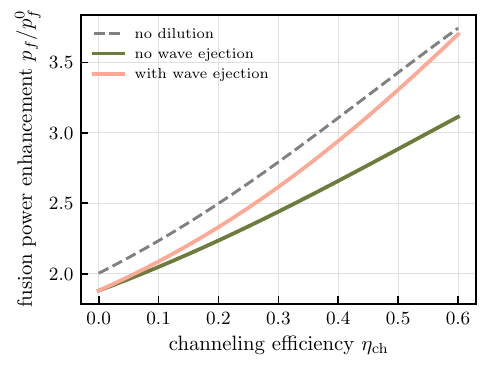}\caption{}\label{fig:ash:b}\end{subfigure}
\caption{(a) Total core helium fraction $n_\mathrm{He}/n_e$ versus $\eta_{\rm ch}$ with and without wave ejection. (b) Fusion power enhancement with dilution feedback. We use $A_J=1.5$ for vector-aligned fuel. Each curve is normalized to its own unpolarized case. The dashed curve shows the dilution-free result of \Cref{fig:gain}.}
\label{fig:ash}
\end{figure}

Because the wave moves each alpha outward as it extracts energy, following $dP_\phi/dE=n_\phi/\omega$, we set $F_\mathrm{ej}=\eta_{\rm ch}$, an assumption consistent with the two-wave Monte Carlo of \Cref{app:com} ($45\%$ of births ejected against $42\%$ of the power diverted) and likely optimistic at low $\eta_{\rm ch}$. We also assume the ejected alphas are extracted at the divertor rather than implanted in the first wall. Without ejection the extra helium takes back part of the channeling enhancement, $3.4$ falling to $2.9$ at $\eta_{\rm ch}=0.5$ as the core helium reaches $3.4\%$; with it the enhancement recovers to $3.3$, near its dilution-free value, and the core helium falls to $1.8\%$ at the same pumping (\Cref{fig:ash}). The ejected alphas reach the edge with the energy channeling did not extract, about $1.6$ MeV each. This energy never heats the plasma, but \Cref{eq:Ti,eq:Te} still treat the full alpha power as deposited: when we scale the deposited alpha power by $1-F_\mathrm{ej}(1.6/3.5)$ the $\eta_{\rm ch}=0.5$ and $0.8$ enhancements in fusion power fall from $3.3$ and $4.5$ to $2.5$ and $2.6$. We leave a treatment that couples $F_\mathrm{ej}$ and $\eta_{\rm ch}$ self-consistently to future work; the ash-model enhancements above should be read with all of these caveats.

Following~\cite{whyte2023,parisi2024}, the core inventory defines the helium particle confinement time
\begin{equation}
\tau^*_\mathrm{He}=\frac{n_\mathrm{He}}{\dot n_\alpha} = (\tau_\mathrm{ash}+\tau_\mathrm{sd}) (1-\eta_{\rm ch}),
\end{equation}
with $\tau^*_\mathrm{He}/\tau_E\approx4$ demonstrated~\cite{sakasai1999} and $10$ a practical limit~\cite{reiter1990}. $\tau_\mathrm{ash}$ is a transport input, the residence time per thermalized ion; $\tau^*_\mathrm{He}$ is an outcome, the average core residence time per alpha born. Channeling shortens the time an alpha spends in the core rather than keeping it out. The ejected alpha fraction leaves within the brief extraction phase, so the average falls by $(1-\eta_{\rm ch})$. Ejection moves the helium inventory from the core to the divertor without changing the flux the pump must handle (at fixed $\dot n_\alpha$).

Another important quantity is the helium enrichment, the ratio of the divertor and core ash concentrations, the ash being the population that cross-field transport delivers to the divertor,
\begin{equation}
\eta_{\mathrm{He}} \equiv \frac{f_{\mathrm{He},\mathrm{div}}}{f_{\alpha,\mathrm{co}}},
\label{eq:etaHe}
\end{equation}
with the core ash-to-tritium ratio
\begin{equation}
f_{\alpha,\mathrm{co}} \equiv \frac{n_\mathrm{ash}}{n_{\mathrm{T},\mathrm{co}}} = \frac{\dot n_\alpha \tau_\mathrm{ash} (1-\eta_{\rm ch})}{n_{\mathrm{T},\mathrm{co}}}. 
\label{eq:falphaco}
\end{equation}
The enrichment is defined using the helium ash rather than the total core helium because only the ash participates in the cross-field transport the enrichment describes.
The divertor ash-to-tritium ratio is
\begin{equation}
f_{\mathrm{He},\mathrm{div}} \equiv \frac{n_{\mathrm{He},\mathrm{div}}}{n_{\mathrm{T},\mathrm{div}}},
\label{eq:ashtofuel}
\end{equation}
where $n_{\mathrm{He},\mathrm{div}}$ and $n_{\mathrm{T},\mathrm{div}}$ are the helium-4 and tritium divertor densities, and we assume there are no fast alphas in the divertor.

In steady-state the divertor pump must exhaust every alpha born, $S_\mathrm{He}\,n_{\mathrm{He},\mathrm{div}}=\dot N_\alpha$, with $\dot{N}_\alpha$ the total alpha birth rate in the plasma. So $n_{\mathrm{He},\mathrm{div}}$ is set by the birth rate and the pump speed, whether the wave or cross-field transport moved the alphas out. The tritium density $n_{\mathrm{T},\mathrm{div}}$ is also unaffected by $\eta_\mathrm{ch}$, which ejects alphas but not fuel. Therefore the divertor helium fraction $f_{\mathrm{He},\mathrm{div}}$ is independent of $\eta_\mathrm{ch}$ at fixed $\dot N_\alpha$, and channeling enters it only through the temperature feedback on the fusion reactivity. Writing the no-ejection condition ($\eta_\mathrm{ch} = 0$) divertor-to-core ratio as the nominal transport enrichment $\eta_\mathrm{He}^{(0)}$ gives the enrichment's dependence on $\eta_\mathrm{ch}$,
\begin{equation}
\eta_\mathrm{He}=\frac{f_{\mathrm{He},\mathrm{div}}}{f_{\alpha,\mathrm{co}}}
=\frac{\eta_\mathrm{He}^{(0)}\,\dot n_\alpha\tau_\mathrm{ash}/n_{\mathrm{T},\mathrm{co}}}{\dot n_\alpha\tau_\mathrm{ash}(1-\eta_{\rm ch})/n_{\mathrm{T},\mathrm{co}}}
=\frac{\eta_\mathrm{He}^{(0)}}{1-\eta_{\rm ch}}.
\label{eq:etaHeboost}
\end{equation}
We set $\eta_\mathrm{He}^{(0)}=1$ in our following analysis. The tritium burn efficiency is~\cite{whyte2023,parisi2024}
\begin{equation}
\mathrm{TBE} \equiv \frac{\dot{N}_\mathrm{T,burn} }{\dot{N}_\mathrm{T,in}}  = \frac{f_{\mathrm{He},\mathrm{div}}\Sigma_\mathrm{HeT}}{1+f_{\mathrm{He},\mathrm{div}}\Sigma_\mathrm{HeT}},
\label{eq:TBE}
\end{equation}
where $\dot{N}_\mathrm{T,burn}$ and $\dot{N}_\mathrm{T,in}$ are the total tritium burn and injection rates for the plasma and
\begin{equation}
\Sigma_\mathrm{HeT}=\frac{S_\mathrm{He}}{S_\mathrm{T}},
\end{equation}
is the helium-to-tritium pumping speed ratio, where $S_\mathrm{He}$ and $S_\mathrm{T}$ are the volumetric pumping rates for helium-4 and tritium in the divertor. We assume $\Sigma_\mathrm{HeT}=0.1$ at standard un-optimized divertor fuel pumping. 

\begin{figure}[tbp]
\includegraphics[width=0.95\columnwidth]{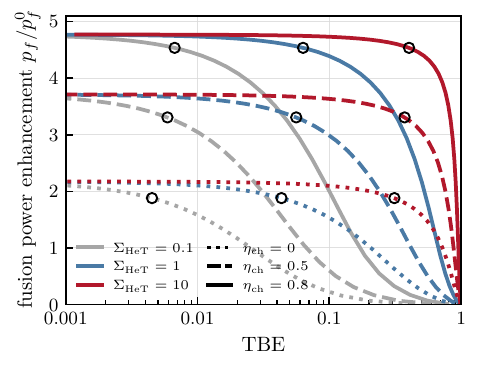}
\caption{Fusion power enhancement versus TBE as the core helium fraction is scanned at fixed pumping ratio, for helium-to-tritium pumping speed ratios $\Sigma_\mathrm{HeT}=0.1$ (standard), $1$, and $10$ (color), and vector-aligned fuel at $\eta_{\rm ch}=0$ (dotted), $0.5$ (dashed), and $0.8$ (solid). Circles mark the self-consistent operating points, where the scanned helium fraction equals the self-consistent value from \Cref{eq:nashss}.}
\label{fig:pump}
\end{figure}

In \Cref{fig:pump} we scan the core helium fraction at fixed pumping ratio. Raising the core helium raises the divertor helium, which raises the burn efficiency (\Cref{eq:TBE}), but it also dilutes the fuel and lowers the fusion power. Each curve therefore trades fusion power against burn efficiency, as first shown in~\cite{whyte2023}. Channeling reduces this tradeoff between fusion power and TBE in two ways: at higher $\eta_{\rm ch}$ we start from a higher fusion power, and channeling ejects some of the helium to the divertor without it residing in the core, so we reach the same TBE with lower helium dilution. The strongly channeled curves in \Cref{fig:pump} are therefore both higher and flatter. \Cref{fig:pump} satisfies the power-reduction relation of~\cite{whyte2023} to within the fast-alpha dilution, with the dilution $(1-2f_\mathrm{dil})^2$, where the helium-to-electron core density ratio is
\begin{equation}
f_{\mathrm{dil}} \equiv \frac{n_\mathrm{He}}{n_e},
\label{eq:dil1}
\end{equation}
also known as the ash dilution fraction \cite{whyte2023}. The fusion power enhancement/degradation is
\begin{equation}
\frac{p_f}{p_f^0} = (1-2 f_\mathrm{dil})^2\, \mathcal{N} A_J,
\label{eq:pDeltaform0}
\end{equation}
with the $(1-2 f_\mathrm{dil})^2$ factor the fuel that the helium displaces through quasineutrality. Here 
\begin{equation}
\mathcal{N}\equiv  \frac{\langle\sigma v\rangle(T_i)}{\langle\sigma v\rangle(T_i^0)},
\end{equation}
is the reactivity ratio of the actual burn point to the helium-free unpolarized, unchanneled reference at the same $n_e$, which is also the reference $p_f^0$, so the dilution enters \Cref{eq:pDeltaform0} only through the explicit $(1-2f_\mathrm{dil})^2$ factor. 

The heat transport model gives $\mathcal{N}A_J=2.0$, $3.4$, and $4.0$ at $\eta_{\rm ch}=0$, $0.5$, and $0.7$ in the small-dilution limit (\Cref{tab:ign}, \Cref{app:arc}). The transport-set helium inventory relates the dilution and the power. Alphas are made at the volumetric rate $\dot n_\alpha=p_f/E_\mathrm{DT}$ of \Cref{eq:ndotalpha}, a fraction $1-\eta_{\rm ch}$ of them stay in the core for a time $\tau_\mathrm{ash}+\tau_\mathrm{sd}$, and the accumulated helium displaces fuel,
\begin{equation}
f_\mathrm{dil}=\frac{p_f}{E_\mathrm{DT}\,n_e}\,(1-\eta_{\rm ch})\,(\tau_\mathrm{ash}+\tau_\mathrm{sd}).
\label{eq:fdilclose}
\end{equation}
Inserting this into \Cref{eq:pDeltaform0} gives,
\begin{equation}
\frac{p_f}{p_f^0}=\left[1-\frac{2\,p_f\,(1-\eta_{\rm ch})(\tau_\mathrm{ash}+\tau_\mathrm{sd})}{E_\mathrm{DT}\,n_e}\right]^2\mathcal{N}A_J,
\label{eq:pDeltaFej}
\end{equation}
with solutions
\begin{equation}
\begin{aligned}
p_f&=\mathcal{N}A_J\,p_f^0\;\frac{1+2s-\sqrt{1+4s}}{2s^2},\\
s&\equiv\frac{2\,\mathcal{N}A_J\,p_f^0\,(1-\eta_{\rm ch})(\tau_\mathrm{ash}+\tau_\mathrm{sd})}{E_\mathrm{DT}\,n_e},
\end{aligned}
\label{eq:pfroot}
\end{equation}
where $s$ is the fuel fraction the dilution-free power would displace. For small $s$ the enhancement is $\mathcal{N}A_J(1-2s)$, so channeling raises the power in two ways: ejection lowers $s$ by $1-\eta_{\rm ch}$, and the hotter ions raise $\mathcal{N}$.

\begin{figure}[tbp]
\includegraphics[width=0.95\columnwidth]{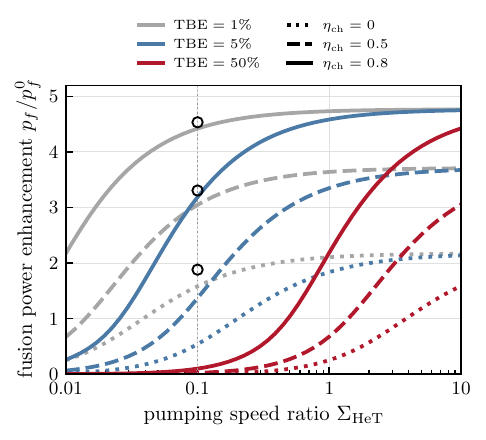}
\caption{Fusion power enhancement versus the helium-to-tritium pumping speed ratio at fixed tritium burn efficiency $\mathrm{TBE}=1$, $5$, and $50\%$ (color), for vector-aligned fuel ($A_J=3/2$) at $\eta_{\rm ch}=0$ (dotted), $0.5$ (dashed), and $0.8$ (solid). The core ash fraction follows from $f_{\alpha,\mathrm{co}}=f_{\mathrm{He},\mathrm{div}}(1-F_\mathrm{ej})$ with $f_{\mathrm{He},\mathrm{div}}=\mathrm{TBE}/[\Sigma_\mathrm{HeT}(1-\mathrm{TBE})]$, at unit divertor enrichment and $F_\mathrm{ej}=\eta_{\rm ch}$. The vertical dotted line marks the standard $\Sigma_\mathrm{HeT}=0.1$, and circles mark the self-consistent operating points of \Cref{fig:ash} there.}
\label{fig:pumpsigma}
\end{figure}

\Cref{fig:pump} shows the fusion power versus TBE, giving the well-known tradeoff~\cite{whyte2023}; this work shows that $\eta_\mathrm{ch}$ is an additional variable allowing simultaneous achievement of high fusion power and TBE, especially at higher $\Sigma_\mathrm{HeT}$. Therefore, the tritium-lean designs benefit from both differential pumping and alpha channeling~\cite{whyte2023,parisi2024pump}. We make the same point in \Cref{fig:pumpsigma} scanning in the pumping ratio: we fixed TBE and scan $\Sigma_\mathrm{HeT}$, and each curve collapses at low $\Sigma_\mathrm{HeT}$ (forcing a dilute core). Wave ejection moves the power collapse to much lower $\Sigma_\mathrm{HeT}$, from $\Sigma_\mathrm{HeT}=0.044$ to $0.011$ at $\mathrm{TBE}=1\%$ as $\eta_{\rm ch}$ increases from $0$ to $0.8$.

\section{Hot-ion mode from turbulence stabilization}
\label{sec:hotion}

\begin{figure}[tbp]
\begin{subfigure}{\columnwidth}\includegraphics[width=0.95\linewidth]{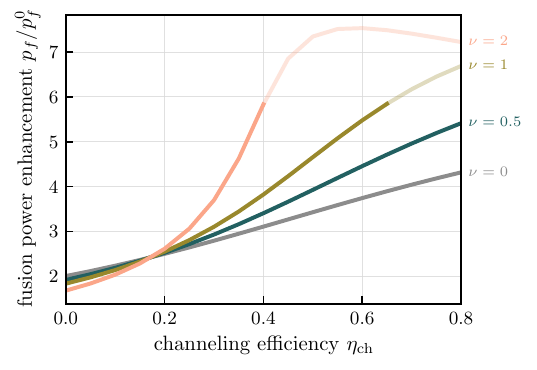}\caption{}\label{fig:hotion:a}\end{subfigure}\\
\begin{subfigure}{\columnwidth}\includegraphics[width=0.95\linewidth]{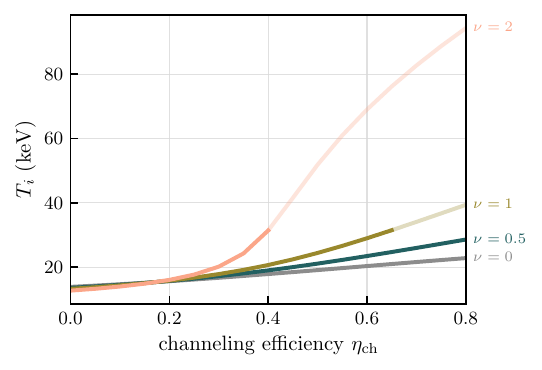}\caption{}\label{fig:hotion:b}\end{subfigure}
\caption{(a) Fusion power enhancement and (b) ion temperature versus channeling efficiency for vector-aligned fuel ($A_J=3/2$), with the ion loss time multiplied by the ITG-feedback factor $H$ of \Cref{eq:hotion} at stiffness $\nu=0$, $0.5$, $1$, and $2$. Curves are faded where the thermal beta at $B=5.3$ T exceeds $10\%$. The $\nu=0$ curves have $H=1$, the burn model of \Cref{app:burn} used throughout \Cref{sec:payoff}.}
\label{fig:hotion}
\end{figure}

The fusion power enhancements of previous sections took the confinement time as given. In this section we let the temperature ratio that channeling produces act back on confinement, through a simple model in which $T_i/T_e$ raises the critical ion temperature gradient. Channeling moves alpha power from the electrons to the ions~\cite{fischherrmann1994}, so the temperature ratio $T_i/T_e$ increases with $\eta_{\rm ch}$ (\Cref{fig:gain}(b)). So far that ratio has been an output. However, $T_i/T_e$ can also feed back on confinement: the critical gradient of the ion-temperature-gradient (ITG) mode \cite{Rudakov1961,Nordman1990,Cowley1991,Nunami2011}, the instability that usually sets ion heat transport, rises with the temperature ratio, $R/L_{T_i}^\mathrm{crit}\propto 1+T_i/T_e$~\cite{romanelli1989,jenko2001}, a scaling derived in the flat-density, adiabatic-electron limit; at strong density gradient the threshold loses much of its $T_i/T_e$ dependence~\cite{casati2008}. In stiff heat transport the achievable ion temperature follows $1 + T_i/T_e$. We model this effect in our transport model with a multiplier $H$ just on the ion energy confinement time,
\begin{equation}
\tau_{E,i}=H\,\tau_E,\qquad H=\left[\frac{1+T_i/T_e}{1+(T_i/T_e)_0}\right]^{\nu},
\label{eq:hotion}
\end{equation}
with $(T_i/T_e)_0=1.13$ the unpolarized, unchanneled temperature ratio, so $H=1$ there, and $\nu$ a stiffness exponent: $\nu=0$ recovers our previous results. \Cref{fig:hotion} shows the results. Without channeling the feedback slightly penalizes aligned fuel, because the extra unchanneled alpha power lands on the electrons and decreases $T_i/T_e$ to $1.04$, below the unpolarized $1.13$. However, channeling opens the hot-ion branch. With it the virtuous feedback loop reinforces, since hotter ions confine better increasing $T_i$ further: at $\nu=1$ the $\eta_{\rm ch}=0.5$ enhancement rises from $3.4$ to $4.7$ at $T_i=25$ keV and $T_i/T_e=1.6$, reaching $6.7$ at $\eta_{\rm ch}=0.8$ with $T_i\approx39$ keV, and at $\nu=2$ the ions reach $52$ keV at $\eta_{\rm ch}=0.5$ and $94$ keV at $0.8$, where the enhancement rolls over as the reactivity passes its maximum. 

A regime of sustained $T_i/T_e$ is known as `hot-ion mode'~\cite{clarke1980hot,strachan1987,keilhacker1999,koide1993characteristic,kurskiev2022first,kaye2023isotope}, the regime of the TFTR supershots and of JET's record D-T performance, and hot-ion states from non-collisional alpha removal have been found in rotating mirrors~\cite{kolmes2021natural}. The loop is stopped by electron-ion coupling, the reactivity peak, and pressure limits rather than running away. This effect is distinct from, and would add to, the fast-ion turbulence stabilization of~\cite{citrin2013,citrin2015,disiena2018}. We also hold $\eta_{\rm ch}$ fixed while the loop moves the plasma far from the conditions where it was computed: the birth anisotropy itself survives the hotter fuel, falling only from $A_\alpha=2.00$ to $1.96$ at $T_i=90$ keV as thermal motion broadens the emission (the Monte Carlo of \Cref{sec:control}), but the slowing-down time roughly triples and the channeled supra-thermal tail makes the thermal $\langle\sigma v\rangle(T_i)$ conservative. 

Several caveats apply to this modeling. At $B=5.3$ T the model's core thermal beta (ratio of thermal to magnetic energy) is $7\%$ already at the feedback-free point ($\nu=0$, $\eta_{\rm ch}=0.5$), $8.6$ to $12\%$ along $\nu=1$, and $16$ to $25\%$ along $\nu=2$; the volume averages that pressure limits constrain are lower by the pressure peaking $p_0/\langle p\rangle=1+\alpha_n+\alpha_T$, about $2.3$ for the {ARC} design profiles~\cite{Sorbom2015} and $2.5$ to $3$ for typical reactor scenarios, so the $\nu\le1$ points likely remain accessible while the $\nu=2$ branch presses the limits, and the curves of \Cref{fig:hotion} are faded above a core beta of $10\%$. Including the helium balance of \Cref{eq:nHethermal} lowers the ion temperatures by about five percent at $\nu=1$ and by a quarter at the $\nu=2$, $\eta_{\rm ch}=0.5$ point, with total helium below $3\%$. The threshold scaling and stiffness exponent are extrapolated far beyond their validated range, the results also have the caveats of the zero-dimensional transport model, higher pressure and its gradients can destabilize other modes, and holding the electron losses fixed is optimistic, since suppressed ITG typically drive transport through trapped-electron and electron-temperature-gradient modes whose thresholds fall as $T_i/T_e$ rises~\cite{casati2008}, with colder electrons increasing the ion-electron collisional coupling. We therefore interpret \Cref{fig:hotion} as showing that the $T_i/T_e$ produced by channeling is large enough for turbulence stabilization to become a further beneficial effect rather than as an accurate prediction. Higher fidelity modeling with profiles and geometry is beyond this work.

\section{Fusion power in an ARC-class plasma with profiles}\label{sec:arctransport}

\begin{figure*}[p]
\begin{subfigure}{0.48\textwidth}\includegraphics[width=0.97\linewidth]{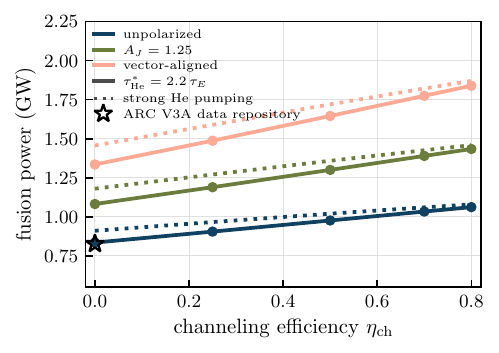}\caption{}\label{fig:arcmain:a}\end{subfigure}\hfill
\begin{subfigure}{0.48\textwidth}\includegraphics[width=0.97\linewidth]{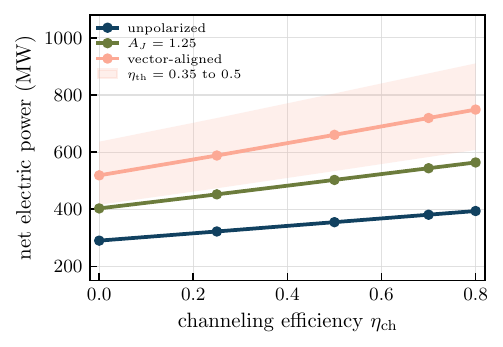}\caption{}\label{fig:arcmain:b}\end{subfigure}\\
\begin{subfigure}{0.48\textwidth}\includegraphics[width=0.97\linewidth]{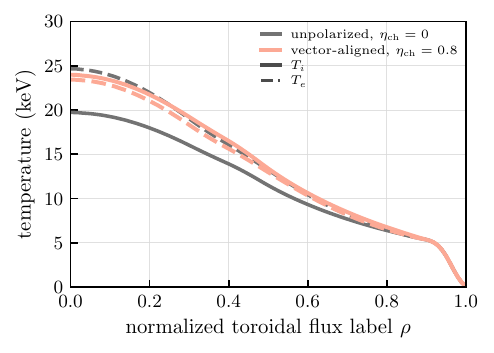}\caption{}\label{fig:arcmain:c}\end{subfigure}\hfill
\begin{subfigure}{0.48\textwidth}\includegraphics[width=0.97\linewidth]{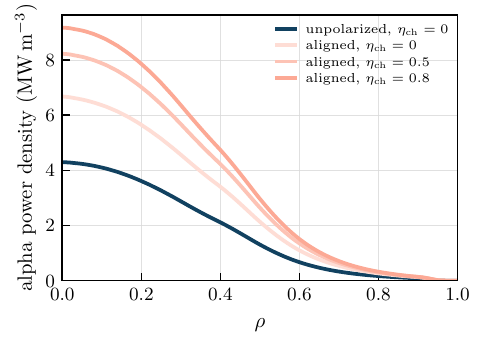}\caption{}\label{fig:arcmain:d}\end{subfigure}\\
\begin{subfigure}{0.48\textwidth}\includegraphics[width=0.97\linewidth]{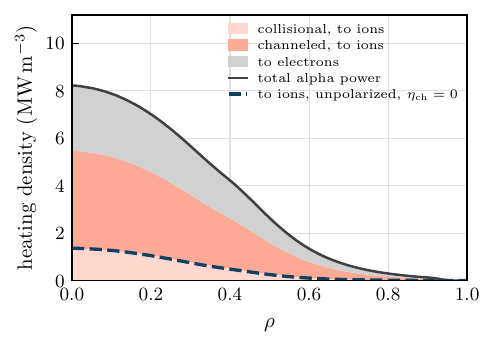}\caption{}\label{fig:arcmain:e}\end{subfigure}\hfill
\begin{subfigure}{0.48\textwidth}\includegraphics[width=0.97\linewidth]{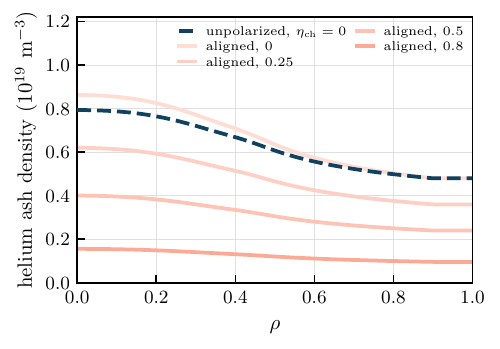}\caption{}\label{fig:arcmain:f}\end{subfigure}
\caption{Transport model of the ARC V3A plasma. (a) Fusion power and (b) net electric power against channeling efficiency, for unpolarized, partially polarized, and fully polarized vector-aligned fuel, using helium confinement $\tau^*_\mathrm{He}=2.2\,\tau_E$ (solid) and with strong helium pumping (dotted, panel a). The star shows the published V3A data-repository state~\cite{hillesheim2026data}. The net electric power uses the V3A plant details in~\cite{hillesheim2026}, the  blanket thermal gain scaled with neutron rate, injected heating, a $42\%$ conversion efficiency, and $100$ MW of recirculating power, with the shaded region varying the electricity conversion efficiency from $0.35$ to $0.5$. (c) Ion (solid) and electron (dashed) temperature profiles for the unpolarized unchanneled plasma and for vector-aligned fuel at $\eta_{\rm ch}=0.8$. (d) Alpha power density. (e) Its deposition at $\eta_{\rm ch}=0.5$, split into ion and electron shares. (f) Thermal helium ash density.}
\label{fig:arcmain}
\end{figure*}

In this section, we calculate the effect of spin-polarized fuel and alpha channeling with profiles on the ARC V3A design~\cite{hillesheim2026,howard2026}, evolving the ion temperature, the electron temperature, and the thermal helium density on the published equilibrium, with the geometry, the sources, and the exchange benchmarked to the published design data~\cite{hillesheim2026data} (\Cref{app:transport}). 

This section is intended to be a plausibility exercise rather than a high-effort modeling effort; we demonstrate that the results of the 0-D modeling in prior sections are not wholly unreasonable. The ARC V3A equilibrium was calculated using sophisticated modeling tools such as POPCON \cite{houlberg1982contour}, TRANSP \cite{pankin2025transp}, ASTRA \cite{Pereverzev2002}, TORAX \cite{citrin2024torax}, CGYRO \cite{Candy2016}, PORTALS \cite{Rodriguez2024a}, and EPED \cite{Snyder2011}. Because such high-fidelity modeling is far beyond the scope of this paper, we use a much simpler set of assumptions that iterate the profiles and equilibrium starting from the published design~\cite{hillesheim2026data}. \Cref{app:transport} describes our model in much more detail, here we cover the main assumptions and results.

Our simplified model uses a critical-gradient-model for the electron and ion temperature profiles. We calculate the critical gradients required to give the profiles of the equilibrium in~\cite{hillesheim2026data}, and assume that those critical gradients remain fixed as we scan the fuel polarization and the alpha channeling efficiency. For the helium density profile, ejection removes the fraction $F_\mathrm{ej}=\eta_{\rm ch}$ of alphas born from the core and lowers the recycling level through the enrichment of \Cref{eq:etaHeboost},
\begin{equation}
n_\mathrm{He}(\rho_b)=(1-F_\mathrm{ej})\,n_\mathrm{He,rec},
\label{eq:nHeprofilearc}
\end{equation}
with $\rho_b$ the boundary radius, and $n_\mathrm{He,rec}$ the recycling level of the unpolarized basecase. We also ensure that the basecase has the reported fuel dilution $(n_\mathrm{D}+n_\mathrm{T})/n_e=0.85$~\cite{howard2026}. Polarization and channeling also enter through the alpha power density
\begin{equation}
p_\alpha=A_J\,n_\mathrm{D}n_\mathrm{T}\langle\sigma v\rangle(T_i) \, \mathcal{E}_0,
\label{eq:palpha}
\end{equation}
where $\mathcal{E}_0=3.5$ MeV and the alpha power into the ions and electrons is
\begin{equation}
\begin{aligned}
& p_{\alpha\to i}=\big[\eta_{\rm ch}+(1-\eta_{\rm ch})f_i\big]\,p_\alpha, \\
& p_{\alpha\to e}=(1-\eta_{\rm ch})(1-f_i)\,p_\alpha.
\end{aligned}
\label{eq:arcsplit}
\end{equation}
We assume that the ejected alphas deposit all of their energy in the plasma, consistent with the designed wave of \Cref{app:com}.

\begin{figure}[tbp]
\begin{subfigure}{\columnwidth}\includegraphics[width=0.95\linewidth]{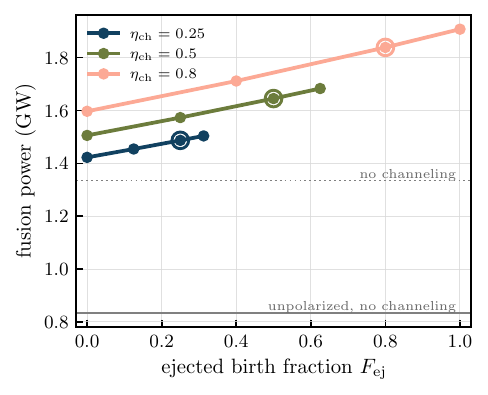}\caption{}\label{fig:arcmech:a}\end{subfigure}\\
\begin{subfigure}{\columnwidth}\includegraphics[width=0.95\linewidth]{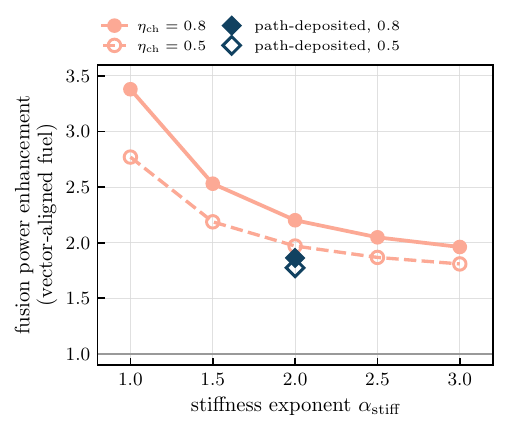}\caption{}\label{fig:arcmech:b}\end{subfigure}
\caption{(a) Fusion power against the ejected birth fraction $F_\mathrm{ej}$ at fixed channeling efficiency, for vector-aligned fuel at the benchmarked helium confinement. Open circles mark our standard choice $F_\mathrm{ej}=\eta_{\rm ch}$ and the horizontal lines mark the unchanneled powers with vector-aligned and with unpolarized fuel. (b) Fusion power enhancement of vector-aligned fuel against the stiffness exponent, at $\eta_{\rm ch}=0.5$ (open) and $0.8$ (filled), with diamonds marking the channeled power deposited along the outward extraction path instead of at the birth radius, at $\alpha_\mathrm{stiff}=2$.}
\label{fig:arcmech}
\end{figure}

\Cref{fig:arcmain} shows the main results, also summarized in \Cref{tab:arccases}. Vector-aligned fuel at $\eta_{\rm ch}=0.8$ raises the fusion power from $0.84$ GW (our `basecase', within $1\%$ of the published data-repository equilibrium~\cite{hillesheim2026data}) to $1.84$ GW, an enhancement of $2.2$ ($2.0$ at $\eta_{\rm ch}=0.5$). The zero-dimensional model of \Cref{sec:payoff} gives power enhancements of $4.5$ and $3.3$ at $\eta_{\rm ch}=0.8$ and $\eta_{\rm ch}=0.5$, the difference being partially reconcilable with the stiffness. In \Cref{fig:arcmech}(b) we show the fusion power multiplier with stiffness, demonstrating much higher power multipliers with less stiff transport. 

With the spin-polarized $\eta_{\rm ch}=0.8$ configuration, the core ions become slightly hotter than the electrons, $T_i = 24.0$ keV, $T_e = 23.4$ keV on axis, whereas the unpolarized plasma is electron-hot, $T_i = 19.7$ keV, $T_e = 24.7$ keV. The on-axis alpha power density rises significantly from $4.3$ to $9.2$ MW\,m$^{-3}$ (\Cref{fig:arcmain}(d)), the ion share of it from $1.4$ to $5.5$ of $8.2$ at $\eta_{\rm ch}=0.5$ (\Cref{fig:arcmain}(e)), and the helium ash falls from $2.75\%$ on axis for aligned unchanneled fuel, above the unpolarized $2.53\%$ since the enhanced reactivity makes ash faster, to $0.50\%$ (\Cref{fig:arcmain}(f)).

\begin{table}[tbp]
\caption{Transport-model ARC V3A cases.}
\label{tab:arccases}
\begin{ruledtabular}
\begin{tabular}{lccc}
 & $P_\mathrm{fus}$ (GW) & $P_\mathrm{e}$ (MW) & on-axis $n_\mathrm{He}/n_e$ \\
\hline
unpolarized, $\eta_{\rm ch}=0$ & $0.84$ & $290$ & $2.5\%$ \\
vector-aligned, $\eta_{\rm ch}=0$ & $1.34$ & $520$ & $2.8\%$ \\
vector-aligned, $\eta_{\rm ch}=0.8$ & $1.84$ & $750$ & $0.5\%$ \\
\end{tabular}
\end{ruledtabular}
\end{table}

In \Cref{fig:arcmech} we plot a scan where $F_\mathrm{ej}$ is not necessarily equal to $\eta_{\rm ch}$. Across stiffness exponents $\alpha_\mathrm{stiff}=1$ to $3$ the power enhancement varies from $3.4$, where the softer gradient response lets the on-axis ion temperature reach $31$ keV, to an enhancement of $2.0$ with stiffer profiles ($\alpha_\mathrm{stiff}=3$). The critical gradients are also held fixed, yet the ion-temperature-gradient threshold rises with $T_i/T_e$~\cite{romanelli1989}, so the hot-ion critical-temperature-gradient-upshift that channeling produces would raise the power enhancement somewhat, although due to the multi-channel nature of heat transport, much higher fidelity modeling is needed to test the response of profiles to changes in $T_i/T_e$.

While we have focused mainly on fusion power increases, the increase in electric power is larger than the fusion one, since the recirculating power is typically fixed while the output grows~\cite{parisi2025electric}. With the V3A plant of~\cite{hillesheim2026} the net electric power increases from $290$ MW at the basecase to $660$ MW at $\eta_{\rm ch}=0.5$ and $750$ MW at $0.8$ (\Cref{fig:arcmain}(b)): the factor $2.2$ in fusion power becomes $2.6$ in electricity. The amplification is relatively modest because the recirculating power, $100$ MW, is small next to the basecase gross electric power, $\eta_\mathrm{th}P_\mathrm{th}=390$ MW before the recirculating power is subtracted. Plants closer to engineering breakeven, where the two are comparable, amplify significantly more~\cite{parisi2025electric}.

\section{Discussion}\label{sec:disc}

We have demonstrated that spin-polarized fuel may enhance fusion power by controlling the velocity-space anisotropy of fusion-born alpha particles. The power enhancement is through the same three mechanisms previewed in \Cref{fig:summary}: the larger alpha source (due to higher cross section), its temperature feedback on reactivity (due to more alpha heating), and the higher channeling efficiency of the perpendicular alpha population, with the wave ejection of ash reducing the core dilution as a further benefit. The transport calculations with ARC-like profiles of \Cref{sec:arctransport} show an enhancement of $2.0$ to $2.2$ under stiff transport and raises the net electric power of an ARC-class plant $2.6$ times~\cite{parisi2025electric}. Given the challenges of tokamaks operating in H-mode~\cite{Kirk2004,Federici2019b,Creely2020,Maingi_2014,Hughes2020,Viezzer2023,Dunne2024,hillesheim2026,howard2026}, spin-polarized fuel and alpha channeling may allow L-mode operation~\cite{Jardin2000,Austin2019,Nelson2023,Paz-Soldan_2024,Wilson2024} to achieve comparable power densities to H-mode with unpolarized, non-channeled fuel.

These benefits rely on several assumptions. First, the anisotropy of fusion-born alphas persists during slowing down over the energy range channeling uses (\Cref{sec:persist}). Second, the alphas must drive waves that then damp on fuel ions through population inversion (\Cref{sec:invert}). Third, the waves amplified by anisotropic alphas must not excessively depolarize the fuel itself: the depolarization rate was demonstrated numerically to be modest for burning plasmas in tokamaks~\cite{cook2026}, although this remains to be demonstrated experimentally~\cite{baylor2023,garcia2023,heidbrink2024,garcia2025}.

While this paper focused on D-T fuel, we also extend our analysis to D-${}^{3}$He in Appendix~\ref{app:dhe3}. Unlike in D-T fuel where 20\% of the fusion energy is released as charged particles, 100\% of the fusion energy from D-${}^{3}$He reactions is released as charged particles, so the wave can act on the full fusion power. Depending on the temperature of the nominal operating point for D-${}^{3}$He, the increase in fusion power could be significantly higher even than for D-T.

Beyond D-${}^{3}$He (Appendix~\ref{app:dhe3}), the same logic may extend to p-${}^{11}$B, where aligning the proton and ${}^{11}$B spins is predicted to raise the rate by up to $60\%$ and to make the proton and alpha emission anisotropic~\cite{dmitriev2006,ahmedweller2014}. Some proton-boron concepts have been suggested to use alpha channeling (with an isotropic birth) such as~\cite{ochs2022}, so polarization-controlled anisotropy may further enhance those schemes.

The birth anisotropy and its persistence depend only on the field direction and collisions, so our results likely apply to many fusion confinement concepts where alpha channeling is beneficial. In stellarators the wave placement and orbit classifications differ significantly by optimization: a quasi-axisymmetric configuration keeps tokamak-like orbits, while a quasi-isodynamic one confines the deeply trapped particles the perpendicular birth preferentially makes, so vector alignment suits it provided the energetic-particle losses stay small~\cite{bader2019,mynick2006}. In mirrors the $\sin^2\theta$ birth deposits alphas away from the loss cone that an isotropic birth feeds, and alpha channeling has been studied there~\cite{zhmoginov2008,fetterman2008}; the same argument extends to other schemes such as the levitated dipole~\cite{simpson2026deuterium}. Finally, the birth anisotropy may have effects in magnetized inertial confinement fusion~\cite{slutz2010,perkins2017}, although alpha channeling is likely too slow to play a major role.

It is important to emphasize that improvements in thermal fusion power usually provide disproportionately large gains in net electric power~\cite{parisi2025electric}; this is because FPPs are expected to devote significant recirculating power to auxiliary systems such as plasma heating and current drive. Therefore, for example, a 50\% increase in fusion power will lead to a much larger increase in net electric power. Furthermore, because many confinement schemes could be power-exhaust-limited, meaning they cannot easily increase their fusion power, an increase in effective fusion reactivity as shown in this work could be used for other benefits such as reduced plasma heating and increased tritium burn efficiency~\cite{whyte2023,parisi2024}, reducing the capital and regulatory costs of a fusion power plant.

Alpha channeling can also drive current. Electron Landau damping absorbs part of the channeled power; launched with an asymmetric $k_\parallel$ spectrum, that absorbed power drives current instead of going to waste, the original use of alpha-amplified waves~\cite{fisch1992,ochsbertelli2015a,ochsbertelli2015b}, and at the $\eta_{\rm ch}=0.5$ operating point it could drive roughly $1$ MA in the ARC power plant (\Cref{app:cd}). Each ejected alpha also hands the wave the toroidal angular momentum $(n_\phi/\omega)\,\Delta E$, available as torque and rotation where the wave damps on ions~\cite{ochsfisch2021,fetterman2008}.

There are important caveats to this work. First, our calculations do not share a single wave channeling scheme: each uses the parameters best-suited to its role, the drive in \Cref{app:fp}, the delivery in \Cref{app:wave}, and the extraction in \Cref{app:cql}. One wave that does all three at once, amplified by the anisotropy, efficient at extraction, and damped on the fuel ions, remains to be designed. Alpha channeling has not yet been demonstrated experimentally, the channeling efficiency $\eta_{\rm ch}$ is tested against reduced solvers (\Cref{app:wave,app:cql}) rather than derived with self-consistent transport, and practical considerations of wave accessibility from the antenna to the core are not addressed. Our zero-dimensional transport model is heuristic and operating-point dependent, and much higher fidelity modeling is required to validate our transport calculations of the ARC-class equilibrium with radial profiles. Our velocity-space calculations are also local, while trapped alphas sample pitch along their bounce orbits, so a treatment with more realistic geometry is required to accurately calculate the inversion boundaries. The same anisotropy that feeds channeling can drive deleterious energetic-particle modes, so it must be used deliberately, and the tensor mode can reduce that drive by depolarization~\cite{cook2026}. Our treatment also omits the wider fast-particle physics of a burning plasma, in which energetic alphas drive or suppress turbulence~\cite{citrin2013,citrin2015,disiena2018} and undergo Alfv\'enic and ripple transport in tokamaks~\cite{gorelenkov2014} and stellarators~\cite{bader2019,mynick2006}, all reshaping the distribution channeling exploits.

None of these caveats undermine the central result, that the polarization state determines the alpha velocity-space distribution, and that anisotropy may be used for enhanced channeling.

Finally, there may be further unrealized capabilities enabled by spin-polarized fuel such as spin-polarized neutrons \cite{schwinger1948polarization,simon1953theory,kulsrud1982,ankner1999polarized}, fusion power burn control~\cite{anderson1993studies,di2025burn,parisi2025doubling,klepper2025feasibility}, and enhanced direct energy conversion~\cite{moir1973venetian} from anisotropic fusion products.

\begin{acknowledgments}
We are grateful for conversations with A. Rutkowski.
\end{acknowledgments}

\section*{Data availability}

The scripts that produce the results in this paper are available from the corresponding author on reasonable request, and will also be made publicly available in a permanent online repository upon publication.

\appendix
\crefalias{section}{appendix}
\section{Quasilinear Fokker-Planck model}\label{app:fp}

We test the persistence and free-energy estimates with a kinetic solver. To do this, we solve for the steady-state alpha distribution $f(v,\xi)$. Here $v$ is the speed in units of the birth speed $v_0$, and $\xi$ is the pitch. The distribution obeys
\begin{equation}
\frac{1}{v^2}\frac{\partial}{\partial v}\big(\nu_s v^3 f\big)+\frac{\nu_d}{2}\frac{\partial}{\partial\xi}\Big[(1-\xi^2)\frac{\partial f}{\partial\xi}\Big]+S=0,
\label{eq:fp}
\end{equation}
which balances three terms. The first term is collisional slowing-down. The second is Lorentz pitch-angle scattering. The third is the birth source $S(v,\xi)$. The source is a narrow shell at $v=v_0$. Its angular shape is $f(\xi)=W(\xi)/A_J$, taken from the chosen mode \Cref{eq:W}. The frequencies $\nu_s$ and $\nu_d$ are the test-particle alpha rates of \Cref{sec:persist}~\cite{nrl}. A sink at low $v$ removes thermalized alphas.

The persistence integral, \Cref{eq:persist}, follows from \Cref{eq:fp} in two steps. Legendre polynomials are eigenfunctions of the pitch-angle term, $\partial_\xi[(1-\xi^2)\,\partial_\xi P_\ell]=-\ell(\ell+1)P_\ell$, so each Legendre moment of $f$ decays at the rate $\ell(\ell+1)\,\nu_d/2$, and the anisotropy moment $\langle P_2\rangle_\theta$ decays at $3\nu_d$. Drag changes the speed but not the pitch, so along the slowing-down history $dv/dt=-\nu_s v$ converts time to speed, $dt=-dv/(\nu_s v)$, and $d\ln\langle P_2\rangle_\theta=3\,\nu_d\,dv/(\nu_s v)$. Integrating from the birth speed down to $v$ gives \Cref{eq:persist}. The $\sin^2\theta$ birth shape has no $\ell=1$ component, and pitch-angle scattering conserves the shell population, so the $\ell=2$ moment describes the whole anisotropy.

\begin{figure}[tbp]
\begin{subfigure}{\columnwidth}\includegraphics[width=0.95\linewidth]{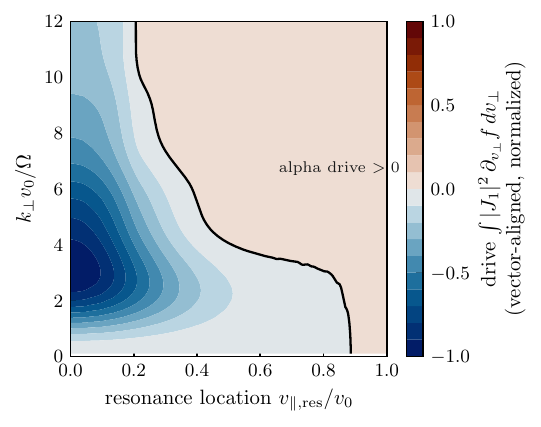}\caption{}\label{fig:drive:a}\end{subfigure}\\
\begin{subfigure}{\columnwidth}\includegraphics[width=0.95\linewidth]{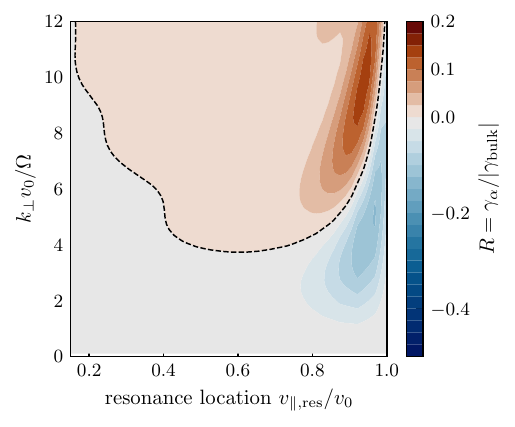}\caption{}\label{fig:drive:b}\end{subfigure}
\caption{(a) Kennel-Engelmann drive $\int |J_1(k_\perp v_\perp/\Omega)|^2\,(\partial f/\partial v_\perp)\,dv_\perp$ on the vector-aligned population, with $f$ the alpha distribution, $k_\perp$ the perpendicular wavenumber, $\Omega$ the alpha cyclotron frequency, and $J_1$ the first-order Bessel function, normalized to its largest magnitude, over resonance location and perpendicular wavenumber. The alpha contribution is positive inside the solid contour. (b) The same drive against bulk damping, the ratio $R=\gamma_\alpha/|\gamma_\mathrm{bulk}|$ of the alpha drive $\gamma_\alpha$ to the bulk damping rate $\gamma_\mathrm{bulk}$, at wave frequency $\omega=0.95\,\Omega$ and alpha density $n_\alpha/n_e=0.5\%$.}
\label{fig:drive}
\end{figure}

The same solver bounds anomalous pitch scattering. Multiplying $\nu_d$ by $1+R$ at unchanged drag, since anomalous processes scatter pitch and position but do not slow the alphas, the anisotropy above $1$ MeV falls from $1.9$ to $1.4$ only at $R\approx13$, the single-wave drive region of \Cref{fig:drive:a} stays open until $R\approx25$, and the inverted region survives to $R\approx30$, though the drive amplitude within that region halves by $R\approx1.5$, and both thresholds are calculated before any wave is applied. The tolerance is large for the same reason the anisotropy persists classically, since above $E_c$ the alpha spends little time against a weak deflection rate, and impurity scattering at $Z_\mathrm{eff}>1$ corresponds to $R\lesssim1$.

The inversion boundaries quoted in \Cref{sec:invert} follow from the drag-dominated limit of \Cref{eq:fp}. Drag conserves pitch. Below the birth shell, the flux in speed is constant. Together these give the factorized steady state $f=C\,W(\xi)/(v^3+v_c^3)$. Here $C$ is a constant. For the vector-aligned mode, $W\propto1-\xi^2$. We differentiate at fixed $v_\parallel$, using $\partial\xi/\partial v_\perp=-\xi v_\perp/v^2$ and $\partial v/\partial v_\perp=v_\perp/v$. We get
\begin{equation}
\frac{\partial f}{\partial v_\perp}=\frac{C\,v_\perp}{v^2\,(v^3+v_c^3)}\left[2\xi^2-\frac{3v^3\,(1-\xi^2)}{v^3+v_c^3}\right],
\label{eq:invcrit}
\end{equation}
which is positive where $2\xi^2/(1-\xi^2)>3v^3/(v^3+v_c^3)$. The first term in the bracket is a gain. It comes from climbing the $\sin^2\theta$ birth weight as the pitch angle grows. The second term is a loss. It comes from descending the slowing-down slope as the speed grows. At the birth speed, with $E_c=0.68$ MeV, the balance sits at $|\xi|=0.76$. The same steps with $W\propto1+3\xi^2$ give the tensor criterion, $\partial f/\partial|v_\parallel|>0$ where $6(1-\xi^2)/(1+3\xi^2)>3v^3/(v^3+v_c^3)$. At the birth speed this gives $|\xi|<0.48$. Pitch-angle scattering blurs these boundaries. The full solver confirms they survive it (\Cref{fig:inv}).

We discretize on a $180\times180$ $(v,\xi)$ grid and solve the linear system directly. The steady state reproduces the analytic slowing-down result (\Cref{fig:fp}(b)). It also gives the perpendicular free-energy fraction of the energetic ($E>1$ MeV) population,
\begin{equation}
\frac{U_\perp}{U}=\frac{\int(\tfrac12 v_\perp^2-v_\parallel^2)\,f\,d^3v}{\int\tfrac12 v^2\,f\,d^3v},
\label{eq:ufree}
\end{equation}
the drive available to a perpendicular wave. We get $U_\perp/U=+0.38$ for vector-aligned, $0$ for unpolarized, and $-0.38$ for tensor (\Cref{fig:fp}(c)). We also add a broadband perpendicular quasilinear diffusion that mimics a channeling wave. It net-damps for every mode. So extracting the bulk alpha energy needs the resonant, energy-space scheme, not a broadband wave.

A wave gains energy from the inversion only if its resonant particles are the inverted ones. For a nearly perpendicular wave at the first cyclotron harmonic, the Kennel-Engelmann growth rate~\cite{kennelengelmann1966} is proportional to $\int |J_1(k_\perp v_\perp/\Omega)|^2\,(\partial f/\partial v_\perp)\,dv_\perp$ at the parallel resonance $v_{\parallel,\rm res}=(\omega-\Omega)/k_\parallel$, with $\Omega=Z e B/m_\alpha$ the alpha cyclotron frequency and $J_1$ the first-order Bessel function, the coupling appropriate to ion Bernstein waves. No explicit $v_\perp$ weight appears because the $v_\perp$ of the phase-space element cancels the $\Omega/v_\perp$ of the cyclotron-resonance gradient; the Landau term, which lacks that factor, keeps its $v_\perp$ weight. The neglected parallel-gradient term, proportional to $(\omega-\Omega)\,\partial f/\partial|v_\parallel|$, is destabilizing for $\omega<\Omega$ on a distribution decreasing in $|v_\parallel|$, so this drive is a conservative underestimate. Without the $|J_1|^2$ selectivity the integral is $-f(v_\perp{=}0)\leq0$, the broadband damping of \Cref{app:fp}. With it, the drive is positive for vector-aligned fuel over roughly half of the wave-parameter range shown, at $v_{\parallel,\rm res}\gtrsim0.2\,v_0$, and negative at every point for unpolarized and tensor fuel (\Cref{fig:drive:a}). The polarization state alone decides whether this drive region exists. This drive is calculated from the alpha distribution before any wave is applied; a wave at operating amplitude flattens the gradient that feeds it, pushing the drive toward zero. A positive alpha term does not by itself sustain the wave, and the drive region lies where the distribution is sparse. We therefore read it as lowering the power needed to sustain a wave tuned into it, with the bulk of the channeling proceeding by the resonant radial path of \Cref{app:cql}, and whether that reduction is realized depends on the specific wave's mode structure and damping.

How close does the drive come to sustaining the wave outright? A full answer needs the kinetic dispersion relation, beyond this work, but a reduced estimate, weighting each species by its squared plasma frequency at its own resonances, gives the scalings. At $\omega=0.95\,\Omega$ the wave is damped mainly at the electron Landau resonance, and the aligned-alpha drive at an alpha density $n_\alpha/n_e=0.5\%$, typical of projected burning-plasma cores~\cite{gorelenkov2014,cook2026}, reaches $14\%$ of the damping, growing linearly with $n_\alpha$ and exponentially as $\omega\to\Omega$ (\Cref{fig:drive:b}). The inversion therefore reduces the wave damping by about ten percent at reactor alpha densities. Whether a specific mode is net driven or damped can be addressed for polarized fuel by direct simulation~\cite{cook2026}.

\section{Burn model}\label{app:burn}

Our zero-dimensional heat transport model evolves the ion and electron temperatures $T_i$, $T_e$. The density $n_e$ is fixed. The equations are
\begin{align}
\tfrac32 n_e\dot T_i ={}& P_\alpha[f_i+\eta_{\rm ch}(1-f_i)]+P_{\rm aux}+P_{ei}\nonumber\\
&{}-\tfrac{3 n_e T_i}{2\tau_E},\label{eq:Ti}\\
\tfrac32 n_e\dot T_e ={}& P_\alpha(1-f_i)(1-\eta_{\rm ch})-P_{ei}\nonumber\\
&{}-P_{\rm br}-P_{\rm sy}-\tfrac{3 n_e T_e}{2\tau_E}.\label{eq:Te}
\end{align}
We evaluate the terms for a $50{:}50$ D-T plasma ($n_\mathrm{D}=n_\mathrm{T}=n_e/2$) in SI units. We take temperatures as energies, except in the empirical coefficients $E_c$, $C_B$, and $c_{\rm sy}$, where $T$ is in keV. The alpha heating is $P_\alpha=\tfrac14 n_e^2\langle\sigma v\rangle(T_i)\mathcal{E}_0 A_J$. Here $\mathcal{E}_0=3.5$ MeV, and $\langle\sigma v\rangle$ is the Bosch-Hale D-T reactivity~\cite{boschhale1992}. The ion-heated fraction is the Stix slowing-down result~\cite{stix1972}
\begin{equation}
f_i=\frac{1}{x_0}\int_0^{x_0}\frac{dy}{1+y^{3/2}},\qquad x_0=\frac{\mathcal{E}_0}{E_c},
\end{equation}
with critical energy
\begin{equation}
E_c=14.8\,A_{\rm f}\,T_e\left[\sum_j \frac{n_j}{n_e}\frac{Z_j^2}{A_j}\right]^{2/3},
\end{equation}
where $A_{\rm f}=4$ is the alpha mass number. The sum runs over background ions $j$ of density $n_j$, charge number $Z_j$, and mass number $A_j$. The bracket equals $5/12$ for the $50{:}50$ mixture. This energy-partition convention gives $E_c\approx33\,T_e\approx0.50$ MeV at $15$ keV, below the $0.68$ MeV drag-rate convention of \Cref{sec:persist}, and the persistence results barely change between the two. The collisional ion-electron exchange is
\begin{equation}
\begin{aligned}
P_{ei}&=\frac{3 m_e}{m_i}\,\frac{n_e (T_e-T_i)}{\tau_e},\\
\tau_e&=\frac{12\pi^{3/2}\varepsilon_0^2\sqrt{m_e}\,T_e^{3/2}}{\sqrt2\,n_e e^4\ln\Lambda},
\end{aligned}
\end{equation}
positive when the electrons are hotter. Here $m_e$ is the electron mass. The mean ion mass is $m_i=(m_D+m_T)/2$, built from the deuteron and triton masses $m_D$, $m_T$. Also, $\varepsilon_0$ is the vacuum permittivity, $e$ is the elementary charge, and the Coulomb logarithm is $\ln\Lambda=17$. The radiative losses are bremsstrahlung and reabsorbed synchrotron,
\begin{equation}
P_{\rm br}=C_B n_e^2\sqrt{T_e},\qquad P_{\rm sy}=c_{\rm sy}\,n_e T_e B^2,
\end{equation}
with magnetic field strength $B$, $C_B=5.34\times10^{-37}$, and $c_{\rm sy}=3.7\times10^{-18}$. The last is an effective coefficient. It stands for strong wall reabsorption. The confinement time degrades with heating power $P=P_{\rm aux}+P_\alpha$ as
\begin{equation}
\tau_E=\tau_0\,(P_{\rm ref}/P)^{0.69},
\end{equation}
following the IPB98(y,2) tokamak scaling~\cite{iter1999}. We apply that scaling to a local power density as a heuristic, with $\tau_0=1$ s and the reference power density $P_{\rm ref}=1$ MW\,m$^{-3}$. We set $n_e=1.5\times10^{20}$ m$^{-3}$, $B=5.3$ T, and $P_{\rm aux}=0.5$ MW\,m$^{-3}$. We then integrate \Cref{eq:Ti,eq:Te} to steady state. 

For \Cref{fig:ash} the burn also includes the helium balance of \Cref{eq:nHethermal}, $\dot n_\mathrm{ash}=\dot n_\alpha(1-F_\mathrm{ej})-n_\mathrm{ash}/(2.5\,\tau_E)$ with $\dot n_\alpha=n_\mathrm{D}n_\mathrm{T}\langle\sigma v\rangle A_J$, plus the fast inventory at its quasi-static value, $n_\mathrm{fast}=\dot n_\alpha(1-F_\mathrm{ej})\,\tau_\mathrm{sd}(T_e)$, with $\tau_\mathrm{sd}$ the Spitzer slowing-down time. Both populations displace fuel, $n_\mathrm{D}=n_\mathrm{T}=(n_e-2n_\mathrm{ash}-2n_\mathrm{fast})/2$ at fixed $n_e$, with $n_\mathrm{fast}$ iterated to a fixed point inside the burn. The wave-ejected fraction $F_\mathrm{ej}=\eta_{\rm ch}$ is removed at the edge. We neglect the helium contribution to the thermal inventories. This fits the heuristic level of the model.

The enhancements quoted in \Cref{sec:payoff} are ratios. The polarized and unpolarized powers come from the same model with the same constants, so much of the model uncertainty divides out. We check how much remains with a sensitivity scan. We vary five parameters at once: the confinement degradation exponent from $0.50$ to $0.85$, the auxiliary power from $0.25$ to $1$ MW\,m$^{-3}$, the density from $1$ to $2\times10^{20}$ m$^{-3}$, the synchrotron coefficient $c_{\rm sy}$ by factors of $0.5$ to $2$, and the coefficient in $E_c$ by factors of $0.7$ to $1.4$, which moves $f_i$. We draw $300$ parameter sets uniformly from these ranges and recompute both enhancements for each set, always normalizing to the unpolarized, unchanneled power at the same parameters.

\begin{figure}[tbp]
\begin{subfigure}{\columnwidth}\includegraphics[width=0.95\linewidth]{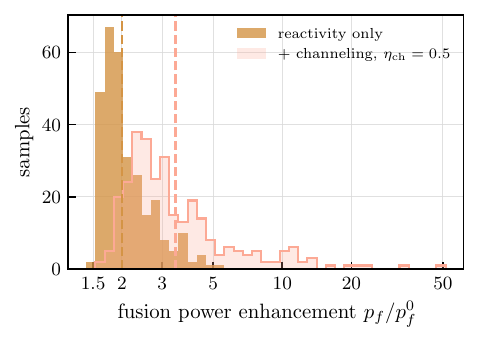}\caption{}\label{fig:sens:a}\end{subfigure}\\
\begin{subfigure}{\columnwidth}\includegraphics[width=0.95\linewidth]{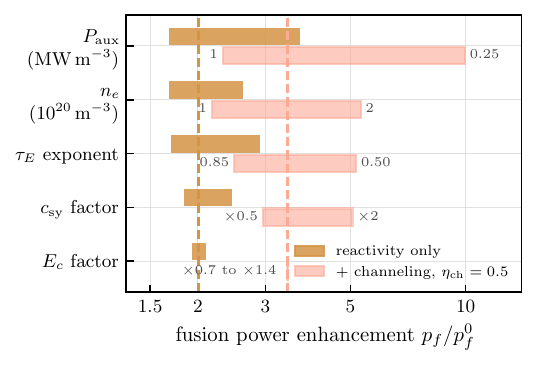}\caption{}\label{fig:sens:b}\end{subfigure}
\caption{Sensitivity of the enhancements to the model parameters of \Cref{app:burn}. (a) Histograms of the fusion power enhancement over $300$ random draws of the five varied parameters, each normalized to the unpolarized, unchanneled power at the same parameters. Dashed lines mark the standard-parameter values, $2.0$ for the reactivity-only case ($A_J=3/2$, $\eta_{\rm ch}=0$) and $3.4$ with channeling ($\eta_{\rm ch}=0.5$). (b) The same two enhancements as each parameter is varied alone across its range, with the others held at their standard values. The small labels give the parameter value at each end of the channeled bar, in the units of the axis labels. Dashed lines show the standard-parameter values as in (a).}
\label{fig:sens}
\end{figure}

\Cref{fig:sens} shows the two distributions. The reactivity-only enhancement has a median of $1.9$ and a $10$ to $90\%$ range of $1.6$ to $2.9$. The enhancement with $\eta_{\rm ch}=0.5$ has a median of $2.9$ and a range of $2.1$ to $7.5$. The distribution is asymmetric. At the most pessimistic corner of the ranges (high auxiliary power, strong degradation, low density, halved synchrotron coefficient) the plasma is firmly driven, the temperature feedback barely closes, and the enhancements fall to $1.5$ and $1.7$ but no further. Where the polarized plasma crosses into a nearly self-heated state and the unpolarized one does not, the ratio instead grows to several tens, the same access effect as in \Cref{fig:dhe3}(b). Where both states self-heat, the ratio returns to about $1.6$.

\Cref{fig:sens}(b) shows the effect of each parameter varied alone, with the others held at their standard values. The auxiliary power is the strongest, moving the channeled enhancement from $10$ to $2.3$ across its range, with the degradation exponent next; the sweep endpoints exceed the $10$ to $90\%$ band because the percentiles exclude the range extremes. The coefficient in $E_c$ barely matters, moving the channeled enhancement from $3.43$ to $3.41$, because whatever fraction collisions fail to deliver to the ions, channeling delivers anyway. The quoted factors of $2.0$ and $3.4$ therefore sit within the $10$ to $90\%$ bands rather than at fine-tuned points.

The five ranges are illustrative choices. The exhaust parameters ($F_\mathrm{ej}$, $\tau_\mathrm{ash}$, and the parasitic wave absorption fraction $\zeta$ of \Cref{app:wave}) are held fixed here; \Cref{fig:ash} and \Cref{fig:pumpsigma} show the consequences of the first directly.

\section{Alpha channeling in D-$^{3}$H\lowercase{e} plasmas}\label{app:dhe3}

The spin physics relevant to the fusion cross section and velocity-space anisotropy of products transfers directly to D-${}^{3}$He from D-T. Helium-3 is spin one-half like the triton, and the fusion reaction occurs through the $3/2^+$ resonance of ${}^{5}$Li (compared with ${}^{5}$He for D-T), so vector alignment increases the cross section by the same factor of up to $3/2$ and \Cref{eq:W} applies, with both products, the $3.6$ MeV alpha and the $14.7$ MeV proton, born $\propto\sin^2\theta$~\cite{kulsrud1986,parisi2025dhe3}. The anisotropy persistence results should apply to the proton as to the alpha, though we compute only the alpha case, and the proton cyclotron frequency is twice the alpha's, so each product can be addressed at its own harmonics.

The channeling effect is larger than in D-T because, apart from D-D side reactions, all of the $18.3$ MeV of fusion energy is born in charged particles, so the wave can act on the full fusion power rather than the alpha fifth. The need is also greater. At a $T_e=60$ keV operating point with equal deuteron and helion densities, the critical energy for the $14.7$ MeV proton is only $0.64$ MeV by the convention of \Cref{app:burn}, so it delivers just $f_i\approx0.09$ of its energy to the ions collisionally, while the $3.6$ MeV alpha, with critical energy $2.6$ MeV at these temperatures, delivers $0.65$. The power-weighted collisional ion fraction is about $0.2$, coincidentally the D-T value, but it now applies to the whole fusion output, and the proton, carrying $80\%$ of that output almost entirely to the electrons, is a good target for channeling.

The burn response should also be more nonlinear than in D-T. The wave reroutes more than five times the power, the rerouting both heats the ions and reduces power into the radiating electrons whose losses are bigger at these temperatures, and the reactivity still climbs with $T_i$ over the relevant range.

\begin{figure}[tbp]
\begin{subfigure}{\columnwidth}\includegraphics[width=0.88\linewidth]{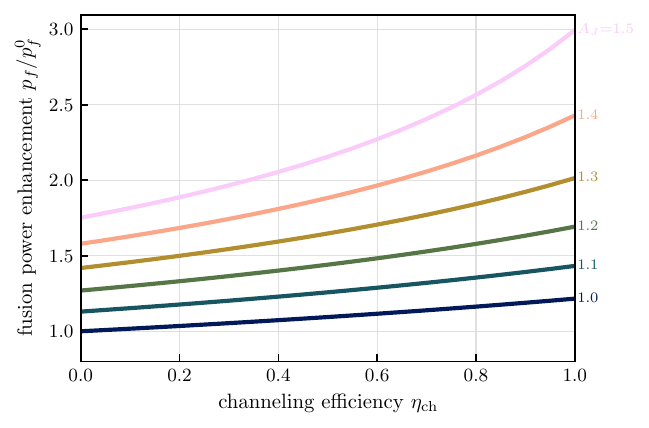}\caption{}\label{fig:dhe3:a}\end{subfigure}\\
\begin{subfigure}{\columnwidth}\includegraphics[width=0.88\linewidth]{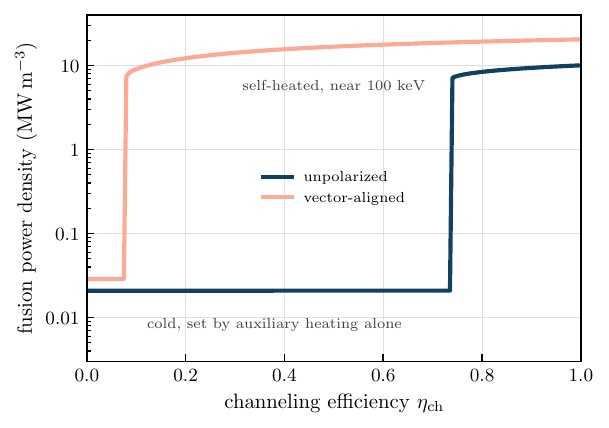}\caption{}\label{fig:dhe3:b}\end{subfigure}\\
\begin{subfigure}{\columnwidth}\includegraphics[width=0.88\linewidth]{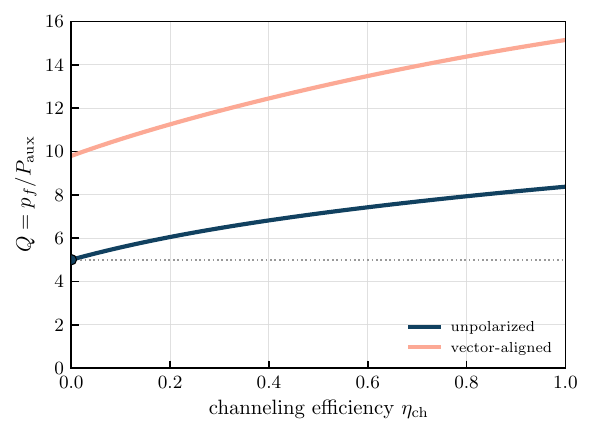}\caption{}\label{fig:dhe3:c}\end{subfigure}
\caption{Channeling in D-${}^{3}$He, with $\eta_{\rm ch}$ acting on the full charged D-${}^{3}$He power and both D-D side channels included (unpolarized, not channeled). (a) Fusion power enhancement versus $\eta_{\rm ch}$ at a driven point ($n_e=1.5\times10^{20}$ m$^{-3}$, $B=7$ T, $\tau_0=3$ s, $P_\mathrm{aux}=1.2$ MW\,m$^{-3}$), normalized to the unpolarized, unchanneled point. (b) Fusion power density at stronger confinement ($n_e=4\times10^{20}$ m$^{-3}$, $B=5$ T, $\tau_0=8$ s, $P_\mathrm{aux}=1$ MW\,m$^{-3}$), from a heated start. (c) $Q=p_f/P_\mathrm{aux}$ versus $\eta_{\rm ch}$, the parameters of (b) with $\tau_0=11$ s and $P_\mathrm{aux}=1.8$ MW\,m$^{-3}$, so the unpolarized, unchanneled plasma (circle, dotted line) sits at $Q=5$.}
\label{fig:dhe3}
\end{figure}

\Cref{fig:dhe3} repeats the enhancement calculation of \Cref{fig:gain} with a D-${}^{3}$He version of the heat transport model where both charged products heat the plasma at their own critical energies and $\eta_{\rm ch}$ acts on the full charged power. The model also includes both D-D side channels at their Bosch-Hale rates. We take the D-D cross section and the D-D product distributions to be unaffected by the polarization, an assumption rather than a result, since the polarization dependence of the D-D channels remains unsettled~\cite{kulsrud1986}. The isotropic D-D products are therefore not channeled and heat the plasma collisionally at their own critical energies, the D-D neutron counts in the fusion output but not in the heating, and secondary reactions of the tritium and helium-3 products are omitted. The side channels carry $3$ to $5\%$ of the fusion power at the self-heated operating points below, and about a third of the far smaller output in the cold states of \Cref{fig:dhe3}(b).

At an operating point with $T_i=36$ keV, the vector-aligned enhancement rises from $1.8$ without channeling to $2.2$ at $\eta_{\rm ch}=0.5$ and $3.0$ at full channeling. The enhancement at this point is smaller than the D-T value even though the wave acts on more than five times the power, because the plasma gain is low: the ratio of fusion power to auxiliary heating is only $0.16$ here versus $5$ for the D-T basecase. Per unit of fusion power the channeling effect is larger than in D-T, taking the ion share from about $0.2$ to $1$ on the whole output, and a self-heated D-${}^{3}$He plasma would see a correspondingly much stronger response.

\begin{figure}[tbp]
\begin{subfigure}{\columnwidth}\includegraphics[width=0.95\linewidth]{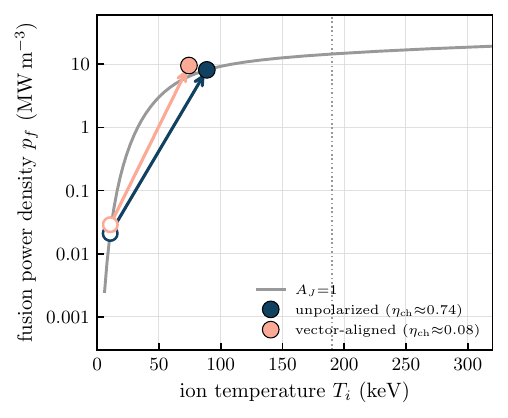}\caption{}\label{fig:dhe3nl:a}\end{subfigure}\\
\begin{subfigure}{\columnwidth}\includegraphics[width=0.95\linewidth]{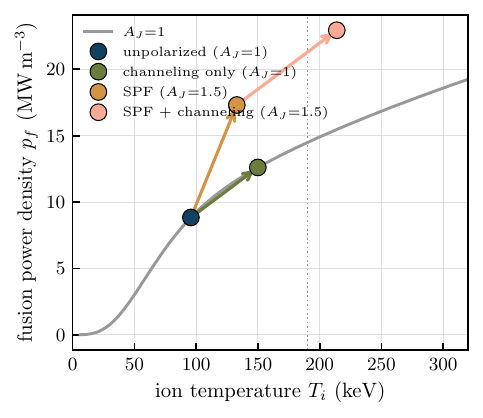}\caption{}\label{fig:dhe3nl:b}\end{subfigure}
\caption{Steady states of \Cref{fig:dhe3} on the reactivity curve, in the style of \Cref{fig:nonlin}(a). In both panels the gray curve is the total fusion power density at $A_J=1$ (D-${}^{3}$He plus D-D) versus $T_i$ at $n_e=4\times10^{20}$ m$^{-3}$, the vector-aligned markers sit above it by the cross-section factor $A_J=3/2$ on the D-${}^{3}$He share (the D-D share is unpolarized), and the vertical dotted line marks the upper end of the fitted reactivity range~\cite{boschhale1992}. (a) The access jump of \Cref{fig:dhe3}(b), on a logarithmic axis, with open markers just below each fuel's channeling threshold and filled markers just above. (b) The four states of \Cref{fig:dhe3}(c), with the channeling markers at $\eta_{\rm ch}=0.5$.}
\label{fig:dhe3nl}
\end{figure}

\Cref{fig:dhe3}(b) shows this stronger response clearly for a case with higher fusion gain. At longer confinement (reference confinement time $\tau_0=8$ s, $n_e=4\times10^{20}$ m$^{-3}$, $B=5$ T, auxiliary heating power $P_\mathrm{aux}=1$ MW\,m$^{-3}$) the model has two roots, a cold driven state and a hot self-heated one, and channeling decides which one the plasma keeps from a heated start. Unpolarized fuel falls to the cold state unless $\eta_{\rm ch}\geq0.74$, while vector-aligned fuel maintains the self-heated state from a value as low as $\eta_{\rm ch}=0.08$, with fusion power rising from about eight to twenty times the auxiliary power as $\eta_{\rm ch}$ grows. Below its threshold the plasma cannot keep itself hot at this auxiliary power and falls to a cold state near $10$ keV, where D-${}^{3}$He fusion is negligible. The flat segments of \Cref{fig:dhe3}(b) are this cold state. Its temperature is set by the auxiliary power alone, so channeling the tiny fusion power basically changes nothing there. At the threshold the channeled ion heating activates the self-heating loop and the plasma jumps to the hot state, at $T_i$ between about $65$ and $90$ keV just above threshold, and the height of the jump is the reactivity ratio between the two temperatures, a factor of several hundred. \Cref{fig:dhe3nl}(a) shows the two states of each fuel on the reactivity curve. The jump is the climb from the cold part of the curve to its flat top, and higher polarization reduces the threshold at which it happens rather than the height of the hot state: the polarization moves the channeling threshold for self-heated D-${}^{3}$He operation by a factor of nine. The hot states reach $T_i$ near $100$ keV and beyond, where our radiation model is least reliable, so we interpret the figure as a statement about access to high-gain D-${}^{3}$He operation.

The sudden jump of \Cref{fig:dhe3}(b) starts from a plasma whose self-heating cannot sustain a high-temperature state without channeling. \Cref{fig:dhe3}(c) instead starts from one that already sustains such a state. We raise the reference confinement time of the same operating point to $\tau_0=11$ s and change the auxiliary power to $1.8$ MW\,m$^{-3}$, so that unpolarized, unchanneled fuel holds a stable self-heated state with $Q=p_f/P_\mathrm{aux}=5$ at $T_i=96$ keV. From this start the response is smooth rather than a jump because there is already appreciable self-heating.

The ion temperature along the aligned curve increases from $133$ keV past $200$ keV, beyond the reliable range of the reactivity fit and of our radiation model, so \Cref{fig:dhe3}(c) should be interpreted with this caveat. Vector alignment alone roughly doubles $Q$ to $10$, since the extra charged power increases through a temperature feedback that is already strong, and channeling raises the aligned plasma to $Q\approx13$ at $\eta_{\rm ch}=0.5$ and $15$ at full channeling. The same channeling without polarization reaches $7$ and $8$, so from a self-heated start the polarization is worth more than full channeling; combining both effects triples $Q$. The response per unit of rerouted power is nonetheless weaker than in D-T, shown in \Cref{fig:dhe3nl}(b), in the style of \Cref{fig:nonlin}(a). The stable $Q=5$ state sits at $T_i=96$ keV, where the logarithmic slope of the reactivity, $d\ln\langle\sigma v\rangle/d\ln T_i$, has fallen to about one, against $2.9$ at the driven point of \Cref{fig:dhe3}(a), so full channeling doubles the ion temperature while the fusion power rises by only two thirds, and the power-degraded confinement returns part of that. A colder $Q=5$ start would sit on the steep part instead, but no stable one exists in this model. The $Q=5$ root at $T_i=55$ keV, which needs $\tau_0\approx14$ s and $P_\mathrm{aux}\approx0.7$ MW\,m$^{-3}$, is thermally unstable and collapses, and scanning the density from $2$ to $6\times10^{20}$ m$^{-3}$ moves the coldest stable $Q\geq5$ state only from about $110$ to about $70$ keV. 

A self-heated D-${}^{3}$He plasma therefore operates near the flat top of its reactivity, where extra ion heating is less valuable. Channeling in D-${}^{3}$He therefore appears to matter most for access to the self-heating regime (\Cref{fig:dhe3}(b)) and for reducing the drive of a cold plasma (\Cref{fig:dhe3}(a)) rather than for boosting an already self-heated plasma.

The exhaust benefit of \Cref{fig:ash} covers both ash species, hydrogen and helium. The main new difficulty is likely confining the large-orbit $14.7$ MeV proton for the duration of its energy extraction. The reactivity and power case for spin-polarized D-${}^{3}$He is made in~\cite{parisi2025dhe3}, and the channeling mechanisms here further increase it.

\section{ARC-class power plant}\label{app:arc}

We now apply the enhancements of \Cref{sec:payoff} to an ARC-class power plant. We use the tritium fuel-cycle model of~\cite{parisi2024} with parameters from~\cite{meschini2023}, applied to an ARC-like machine~\cite{Sorbom2015,hillesheim2026}. We hold $n_e T\tau_E$ fixed and assume $\mathcal{N}A_J=2.0$ for $\eta_\mathrm{ch}=0 $, $\mathcal{N}A_J=3.4$ at $\eta_{\rm ch}=0.5$, and $\mathcal{N}A_J=4.0$ at $\eta_{\rm ch}=0.7$. In this model the plasma gain is $Q=5/(kC-1)$, where $C$ includes the dilution, the polarization, and the tritium-fraction optimization of the burn (see~\cite{parisi2024} for more details), and the ignition boundary is at $kC=1$, with $kC<1$ the ignited region beyond it. We require the unpolarized plasma at the pumping ratio $\Sigma_\mathrm{HeT}=0.3$ to reproduce the ARC V3A design point, $P_f=1130$ MW at $Q=51$~\cite{hillesheim2026}, giving a nominal burn efficiency of $\mathrm{TBE}=0.5\%$. This $\Sigma_\mathrm{HeT}=0.3$ is consistent with the standard $0.1$ of \Cref{sec:payoff}: the exhaust enters through the product $\eta_\mathrm{He}\Sigma_\mathrm{HeT}$, which is $0.19$ here against $0.1$ at unit enrichment there.

\begin{figure*}[p]
\includegraphics[width=0.86\textwidth]{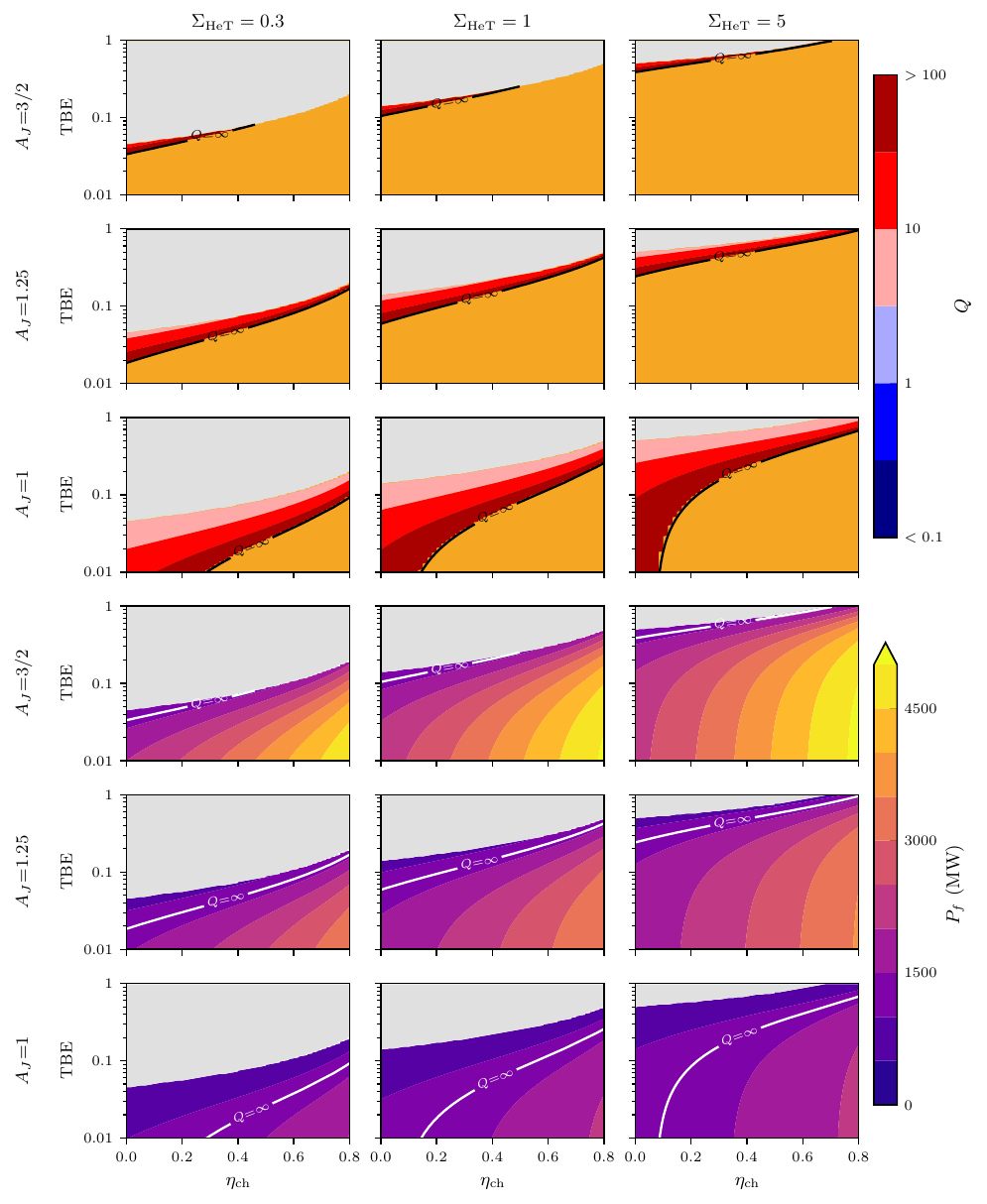}
\caption{Plasma gain $Q$ (top three rows) and fusion power $P_f$ (bottom three rows) versus tritium burn efficiency (TBE) and channeling efficiency $\eta_{\rm ch}$, for pumping ratios $\Sigma_\mathrm{HeT}=0.3$ (nominal), $1$, and $5$ (columns) and cross-section factors $A_J=3/2$, $1.25$, and $1$ (rows within each block), with $\mathcal{N}A_J$ and the enrichment (\Cref{eq:etaHeboost}) computed self-consistently at every point and the model benchmarked to the ARC V3A design point~\cite{hillesheim2026}. Orange regions inside the $Q=\infty$ contour ignite, and grey regions have no steady state solution.}
\label{fig:arc}
\end{figure*}

\Cref{fig:arc} shows the plasma gain and the fusion power against the burn efficiency and the channeling efficiency, with $\mathcal{N}$ calculated self-consistently from the fuel cycle and the 0-D transport model for all the points. Three results stand out. 

First, vector-aligned fuel is ignited across the bottom of every panel, since ignition requires only $\mathcal{N}A_J\approx1.2$ and full alignment gives $2.0$. 

Second, channeling extends the ignited region to higher TBE: at the nominal $\Sigma_\mathrm{HeT}=0.3$, from TBE$=0.033$ unchanneled to TBE$=0.14$ at $\eta_{\rm ch}=0.7$. Much of this comes from alpha ejection raising the enrichment (\Cref{eq:etaHeboost}): freezing the enrichment at $\eta^{(0)}_\mathrm{He}=0.63$ lowers the maximum ignited TBE at $\eta_{\rm ch}=0.7$ from TBE$=0.14$, $0.38$, and $0.95$ to TBE$=0.045$, $0.14$, and $0.49$ at $\Sigma_\mathrm{HeT}=0.3$, $1$, and $5$, so at the plant level the ejection enrichment is the largest single benefit of channeling. 

Third, differential pump capability is essential to handle the extra helium generated from spin-polarized fuel and channeling: at $\Sigma_\mathrm{HeT}=1$ the ignited region reaches $\mathrm{TBE}=0.10$ to $0.38$, and at $\Sigma_\mathrm{HeT}=5$ it reaches $0.38$ to $0.95$, with the fusion power up to $4.8$ GW (at ultra-low tritium burn efficiency).

\section{Wave-kinetic estimate of the delivery efficiency}\label{app:wave}
We test the delivery efficiency $\varepsilon_w$ assumed in \Cref{sec:payoff} with a reduced model of the resonant channeling operator. Each wave kick removes energy and canonical toroidal angular momentum together, so the quasilinear path obeys $dP_\phi/dE=n_\phi/\omega$~\cite{fisch1992,fisch2015}, and an alpha that feeds the wave also moves outward in radius. Along that path the wave acts like an extra drag in speed, and the problem reduces to a one-dimensional kinetic equation for the slowing alpha distribution $f(v)$,
\begin{equation}
\frac{1}{v^2}\frac{\partial}{\partial v}\Big[\big(\nu_s+\nu_w\big)\,v^3 f\Big]-\nu_{\rm ash}(v)\,f+S(v)=0 .
\label{eq:wavekin}
\end{equation}
The first term combines the collisional drag of \Cref{app:fp} (frequency $\nu_s$) with the wave drag $\nu_w$, which acts only in a broadened cyclotron-harmonic band, $\nu_w(v)=\alpha_w\,\nu_{s0}\,g(E)$. Here $g(E)$ is a Gaussian centered at $0.55$ of the birth energy with width $0.30$, $\nu_{s0}$ is the birth slowing rate, and $\alpha_w=D_0/\nu_{s0}$ is the amplitude we scan, with $D_0$ the quasilinear diffusion rate at the band center. The source $S(v)$ is the narrow birth-speed source of \Cref{app:fp}, and the ash sink $\nu_{\rm ash}(v)$ removes thermalized alphas below a cutoff well beneath the band. \Cref{eq:wavekin} is first order in $v$, so it admits one boundary condition: we impose $f=0$ just above the birth speed and integrate downward, with the ash sink $\nu_\mathrm{ash}$ removing the flux below the low-speed cutoff.

\begin{figure*}[tp]
\begin{subfigure}{0.46\textwidth}\includegraphics[width=\linewidth]{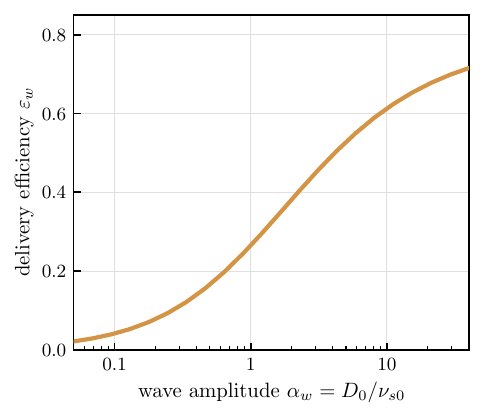}\caption{}\label{fig:wave:a}\end{subfigure}\hfill
\begin{subfigure}{0.46\textwidth}\includegraphics[width=\linewidth]{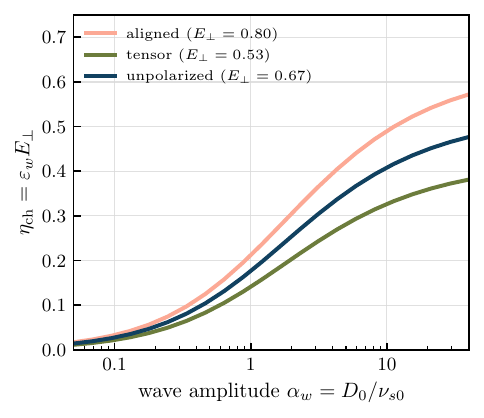}\caption{}\label{fig:wave:b}\end{subfigure}
\caption{Delivery efficiency from the reduced wave-kinetic solver (\Cref{eq:wavekin}). (a) Delivery efficiency $\varepsilon_w$ versus wave amplitude $\alpha_w=D_0/\nu_{s0}$, with $D_0$ the quasilinear diffusion rate at the band center and $\nu_{s0}$ the slowing rate at the birth speed. (b) Channeling efficiency $\eta_{\rm ch}=\varepsilon_w E_\perp$, with $E_\perp$ the perpendicular energy fraction, for the three modes.}
\label{fig:wave}
\end{figure*}

In steady state every alpha born is slowed by collisions, channeled by the wave, or removed as ash, so the birth power divides as $P_\alpha=P_{\rm coll}+P_{\alpha\to w}+P_{\rm ash}$, with $P_{\rm coll}=\int 2\nu_s E f\,d^3v$, $P_{\alpha\to w}=\int 2\nu_w E f\,d^3v$ the power the wave extracts, and $P_{\rm ash}$ the rest. The computed balance closes to within $1\%$. With the extracted fraction $\chi=P_{\alpha\to w}/P_\alpha$, the delivery efficiency is
\begin{equation}
\varepsilon_w=\frac{\chi}{1+\zeta},\qquad \eta_{\rm ch}=\varepsilon_w E_\perp,
\label{eq:epsw}
\end{equation}
with $\zeta=0.25$ the parasitic (electron-Landau) absorption relative to the useful ion deposition and $E_\perp$ the perpendicular fraction of the birth energy. The factorization assumes the perpendicular-resonant single-wave class, whose extraction is almost entirely perpendicular (\Cref{app:cql}). A two-wave scheme can remove the parallel energy as well~\cite{herrmann1997}, and the multi-harmonic or broadband schemes needed for the largest $\eta_{\rm ch}$ relax the $E_\perp$ limit rather than respect it.

\Cref{fig:wave}(a) shows $\varepsilon_w$ rising with wave amplitude, reaching the $0.625$ we use only at strong drive, so the value in \Cref{sec:payoff} sits at the optimistic end of the achievable range. The resulting $\eta_{\rm ch}=\varepsilon_w E_\perp$ reproduces the $0.5$ and $0.3$ taken for the vector-aligned and tensor modes (\Cref{fig:wave}(b)). The model is blind to pitch, so its modes differ only through $E_\perp$; the pitch-resolved solver of \Cref{app:cql} supplies the ranking of the modes and the resonance overlap. The wave is amplified at the alphas' expense and damps on the fuel ions, so in steady state the antenna seeds the wave and covers the parasitic fraction $\zeta$ of the circulating power, which \Cref{eq:epsw} subtracts. Whether the required $k_\perp\rho_\alpha\approx2.5$ spectrum can be launched and propagated to the core is a separate wave-engineering question~\cite{valeofisch1994,ochs2022}. A fully predictive treatment would evolve the wave amplitude self-consistently against the antenna and the equilibrium profiles, which we leave to future work.

\section{Velocity-space calculation}\label{app:cql}

\begin{figure*}[tp]
\begin{subfigure}{0.24\textwidth}\includegraphics[width=\linewidth]{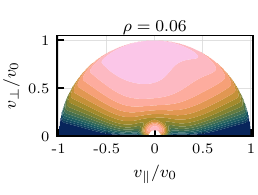}\caption{}\label{fig:radial:a}\end{subfigure}\hfill
\begin{subfigure}{0.24\textwidth}\includegraphics[width=\linewidth]{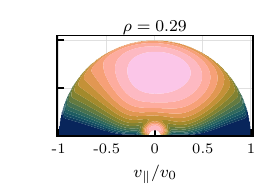}\caption{}\label{fig:radial:b}\end{subfigure}\hfill
\begin{subfigure}{0.24\textwidth}\includegraphics[width=\linewidth]{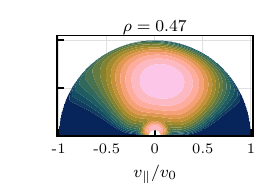}\caption{}\label{fig:radial:c}\end{subfigure}\hfill
\begin{subfigure}{0.24\textwidth}\includegraphics[width=\linewidth]{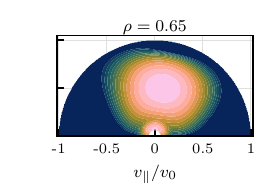}\caption{}\label{fig:radial:d}\end{subfigure}\\
\centerline{\includegraphics[width=0.4\textwidth]{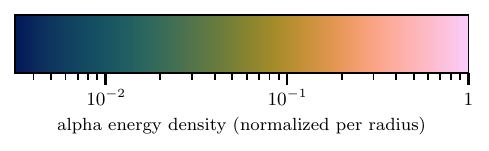}}
\caption{Alpha energy density in velocity space at four minor radii $\rho$, for vector-aligned fuel under a single channeling wave at strong drive, from the velocity-space solver of this appendix with a birth source peaked near the magnetic axis. Each panel is normalized to its own maximum on a logarithmic color scale. The wave's parallel spectrum is directed, resonant near $v_\parallel=0.25\,v_0$ with $v_0$ the birth speed.}
\label{fig:radial}
\end{figure*}

We want the channeling efficiency without assuming the perpendicular energy fraction. So we solve for the alpha distribution $f(v,\xi,\rho)$ in speed, pitch $\xi=v_\parallel/v$, and a normalized minor radius $\rho$ (a proxy for $P_\phi$). We use a quasilinear Fokker-Planck model of the type used in CQL3D~\cite{cql3d}. Collisions are the linearized test-particle operator on a Maxwellian electron, deuterium, and tritium background, with Rosenbluth-potential coefficients. This operator holds a Maxwellian stationary, conserves particles, and reproduces the slowing-down anisotropy of \Cref{sec:persist}. The wave is the resonant Kennel-Engelmann quasilinear operator~\cite{kennelengelmann1966} for a cyclotron-harmonic wave. At the resonance $\omega-k_\parallel v_\parallel-n\Omega=0$, with $n$ the harmonic number, the diffusion is one-dimensional in velocity. It points along each wave kick $(\Delta v_\parallel,\Delta v_\perp)\propto(k_\parallel v_\perp,n\Omega)$. It is weighted by the quasi-electrostatic finite-Larmor-radius coupling $|J_n(k_\perp v_\perp/\Omega)|^2$. We build the operator in a symmetric form, which keeps it numerically stable. The scan of \Cref{fig:cql} uses $n=2$, a resonance at $v_{\parallel,\rm res}=0.25\,v_0$, $k_\parallel\rho_\alpha=0.15$, and $k_\perp\rho_\alpha=2.5$, with $\rho_\alpha$ the birth gyroradius, so $\omega\approx2.0\,\Omega$. We normalize its amplitude to the deflection frequency at the birth speed. That reference rate is much smaller than the slowing rate used in \Cref{app:wave}, so the two amplitude axes are not directly comparable. Removing energy $dE$ moves the alpha radially by $dP_\phi=(n_\phi/\omega)\,dE$. So the wave diffusion includes a radial part. We normalize it so that a full-energy loss moves the alpha across the minor radius. An edge sink removes channeled alphas cold.

Keeping this radial transport turns the velocity-space wave damping of the bare operator into net channeling. We integrate to steady state from the anisotropic birth source. We read $\eta_{\rm ch}$ off the energy balance $P_\alpha=P_{\rm coll}+P_{\alpha\to w}+P_{\rm edge}+P_{\rm ash}$, with $P_{\alpha\to w}$ the power the wave extracts and $P_{\rm edge}$ the power the edge sink removes. The extracted power is more than $99\%$ perpendicular for the vector-aligned and unpolarized modes ($98.7\%$ for tensor). This is an output of the coupling, not an input. The results are grid-converged to a few percent.

The single-wave efficiency saturates near $\eta_{\rm ch}\approx0.09$, because the $|J_n|^2$ coupling weakens as the alpha is channeled to low $v_\perp$. That value is lower than the optimistic reduced estimate of \Cref{app:wave}. We also tested multi-wave spectra, giving each wave its own resonance, coupling, and diffusion path. A second wave at $k_\perp\rho_\alpha=4.5$ raises $\eta_{\rm ch}$ at fixed per-wave amplitude, $0.069$ against $0.038$ at $\alpha_w=20$, but at strong drive it reaches $0.079$ against the single wave's $0.088$, the same saturated value. A third wave at $k_\perp\rho_\alpha=12$ couples to the slowed population and makes the net exchange negative. The saturation is therefore set by the resonant fraction of the population, and perpendicular waves added to reach the rest slide toward the broadband limit of \Cref{app:fp}.

We reran the same solver with five second-harmonic waves at resonant velocities $v_\mathrm{res}/v_0=0.2$ to $0.6$ and find the channeling efficiency factor $A_\eta$ insensitive to the spectrum: $A_\eta$ is $1.44$ and $1.46$ at moderate and strong per-wave drive, against $1.47$ for the single wave, while the five-wave spectrum raises the absolute aligned efficiency from $0.055$ to $0.086$ for an equivalent per-wave amplitude.

Wave amplification by the alphas raises $A_\eta$ further. The alphas born from vector-aligned fuel drive the wave through their population inversion, offsetting the fraction $R$ of its damping (\Cref{fig:drive}(b)), so at fixed launched power the sustained amplitude rises by $1/(1-R)$. Evaluating the aligned efficiency at that amplitude against the unpolarized efficiency at the unboosted one raises $A_\eta$ to $1.65$ and $1.56$ at $R=0.14$, and to $1.91$ and $1.70$ at $R=0.3$, at moderate and strong drive (\Cref{fig:multiplier}). We attribute the gap between these values and the $\eta_{\rm ch}\approx0.5$ to $0.7$ of~\cite{herrmann1997,fischherrmann1994,zhmoginov2008} to three things we deliberately omit from this solver: the $\mu$-conserving second wave, which extracts through the spatial gradient at any $v_\perp$ and so keeps extracting after the perpendicular coupling has died away; the toroidal variation of $\Omega$, which sweeps a single frequency through resonance with essentially every alpha along its orbit; and amplitudes strong enough that wave diffusion beats collisional drag along the whole path.

The two-wave Monte Carlo of \Cref{app:com}, which includes the first two and neglects collisions, diverts $42\%$ of the alpha power with both waves and $60\%$ with the single wave we design for aligned births, so we interpret the two calculations as lower and upper estimates of the achievable efficiency. Adding a $\mu$-conserving Landau-resonant wave at the same radial coupling net-heats the alphas whether the two resonances overlap or are separated, so the tuned two-wave pairing of~\cite{herrmann1997} needs per-wave radial steps and amplitudes beyond this scan. The large diverted fractions of that scheme, which our Monte Carlo approaches (\Cref{app:com}), count the full trajectory of each alpha to the wall at tuned amplitudes rather than a steady-state wave gain; we leave tuning of the continuum two-wave system to future work. The ratio between modes, however, appears to be insensitive to amplitude. It gives the contributions shown in \Cref{fig:cql}(b). \Cref{fig:radial} shows the channeling at work in space. Alphas born near the magnetic axis arrive at larger radius progressively colder, because the wave's quasilinear path ties each outward step to the perpendicular energy it extracts. The mean energy of the population above the ash region falls from $0.33\,\mathcal{E}_0$ near the axis to $0.20\,\mathcal{E}_0$ at $\rho=0.65$, and essentially none of the population beyond mid-radius retains its birth energy.

\section{Guiding-center orbit classification}\label{app:com}

\begin{figure}[tbp]
\begin{subfigure}{\columnwidth}\includegraphics[width=0.95\linewidth]{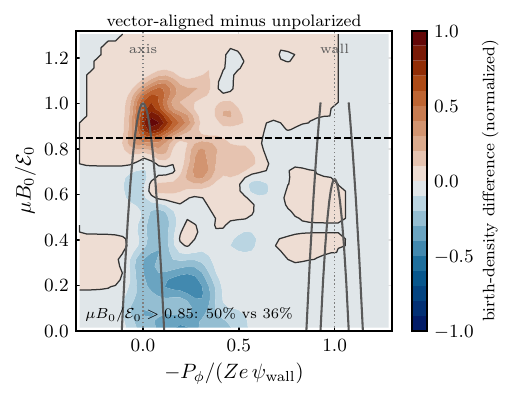}\caption{}\label{fig:comdiff:a}\end{subfigure}\\
\begin{subfigure}{\columnwidth}\includegraphics[width=0.95\linewidth]{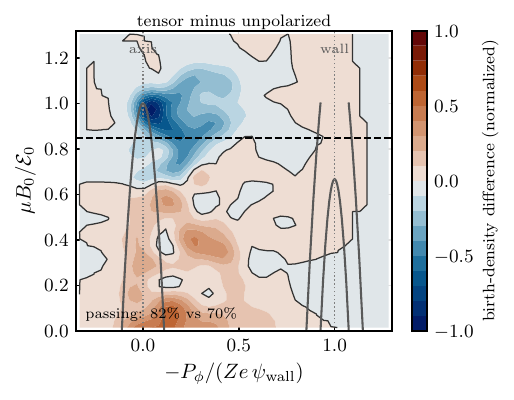}\caption{}\label{fig:comdiff:b}\end{subfigure}
\caption{Birth-density difference in constants-of-motion space, (a) vector-aligned minus unpolarized and (b) tensor minus unpolarized, each birth distribution normalized before subtracting and both panels on one scale. The horizontal coordinate is $-P_\phi/(Ze\,\psi_\mathrm{wall})$, with $P_\phi$ the canonical toroidal angular momentum, $Ze$ the alpha charge, and $\psi_\mathrm{wall}$ the poloidal flux at the wall, approximately $0$ at the axis and $1$ at the wall (dotted lines). It is the flux-surface of the orbit up to the parallel momentum, so co-passing alphas are below $0$ near the axis and counter-passing alphas beyond $1$ near the wall, a range of about one orbit width. Red regions gain births under polarization. The dashed line marks $\mu B_0/\mathcal{E}_0=0.85$, with $\mu$ the magnetic moment and $B_0$ the on-axis field, above which Herrmann and Fisch set the low-frequency wave's $dP_\phi/dE$ three times larger, and the gray curves show the axis and wall-touching orbit boundaries.}
\label{fig:comdiff}
\end{figure}

\begin{figure}[tbp]
\includegraphics[width=0.95\columnwidth]{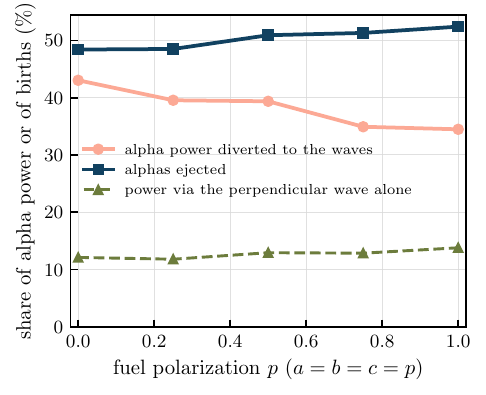}
\caption{Two-wave Monte Carlo versus fuel polarization $p$ ($a=b=c=p$, with $a$ and $b$ the deuteron and triton vector polarizations and $c$ the deuteron tensor polarization), at wave amplitudes tuned for isotropic fuel, showing the alpha power diverted to the waves and the part of it diverted by the perpendicular wave alone, both as shares of the total alpha birth power, and the alphas ejected as a share of all births.}
\label{fig:pscan}
\end{figure}

The polarization changes the orbits on which the alphas are born. Herrmann and Fisch designed their two-wave extraction in the space of magnetic moment $\mu=m_\alpha v_\perp^2/2B$ and canonical toroidal angular momentum $P_\phi$: a high-frequency wave breaks the invariance of $\mu$ and extracts perpendicular energy, acting most strongly near $\mu B_0/\mathcal{E}_0=1$, with $B_0$ the on-axis field, while a low-frequency wave conserves $\mu$ and pushes passing particles outward in $P_\phi$~\cite{herrmann1997}. 

We calculate the births in this space by integrating guiding centers (parallel streaming, mirroring, and the vertical drift, at conserved energy and $\mu$) in a circular large-aspect-ratio tokamak equilibrium with the reactor parameters of Herrmann and Fisch: $R_0=5.4$ m, $a=R_0/3$, $B_0=6$ T, $I_p=16.3$ MA, current density $\propto1-(r/a)^2$ in minor radius $r$, fusion-source profile $\propto(1-(r/a)^2)^4$, birth energy $\mathcal{E}_0=3.5$ MeV, and the pitch distribution of the chosen polarization mode (\Cref{eq:W}). An orbit is trapped if $v_\parallel$ reverses, lost if it reaches $r=a$, and passing otherwise. In $P_\phi=m_\alpha R\,v_\parallel B_\phi/B-Ze\,\psi(r)$, with $B_\phi$ the toroidal field and $\psi$ the poloidal flux, the first term is smaller than the second by roughly the ratio of the alpha poloidal gyroradius to the minor radius, about $0.08$ here, so $P_\phi\approx-Ze\,\psi(r)$ labels the orbit's flux surface, but only to within an orbit width; we therefore work with whole orbits, read from their midplane crossings, rather than point radii. At these parameters no birth orbit is lost for any mode. Vector alignment puts half of the births at $\mu B_0/\mathcal{E}_0>0.85$, against $36\%$ for unpolarized and $21\%$ for tensor fuel, with trapped fractions of $43\%$, $30\%$, and $18\%$; the tensor mode is $82\%$ passing and suits the $\mu$-conserving wave. \Cref{fig:comdiff} shows the relocation as the birth-density difference from unpolarized fuel.

The two-wave Monte Carlo follows~\cite{herrmann1997}. Each alpha takes random energy kicks along one wave's quasilinear path in $(E,\mu,P_\phi)$ with collisions neglected and kicks that would leave no confined orbit rejected. The perpendicular-resonant wave follows $d(\mu B_0)/dE=1$ and $dP_\phi/dE=1.4\,Ze\,\psi_\mathrm{wall}/\mathcal{E}_0$, with $\psi_\mathrm{wall}$ the poloidal flux at the wall, kicks only where the orbit crosses the layer $R>R_0+0.35\,a$, and an interaction probability $\propto\mu B_0/E$ as a schematic finite-Larmor-radius coupling; the $\mu$-conserving wave follows $d\mu=0$ with $dP_\phi/dE=1.6\,Ze\,\psi_\mathrm{wall}/\mathcal{E}_0$, weakened three times below $\mu B_0/E=0.85$ as in the original. Cooling moves each alpha outward in $P_\phi$, so it ends at the wall cooled, and the extracted energy is $\mathcal{E}_0-E_\mathrm{exit}$, with $E_\mathrm{exit}$ its energy at ejection. On unpolarized fuel the two waves together divert $42\%$ of the alpha power, counting the energy extracted from every alpha regardless of whether it reaches the wall; $45\%$ of births are ejected, giving up $1.9$ MeV each and exiting with about $1.6$ MeV. The tuned original ejects $73\%$ and diverts $51\%$~\cite{herrmann1997}, the difference reflecting our untuned relative amplitudes and the finite number of kicks we allow each alpha. The percentages have a statistical uncertainty of about three points at $10^3$ alphas per run. Scanning the polarization $p$ ($a=b=c=p$) at these fixed amplitudes, the diverted power falls from $43\%$ to $35\%$ as $p\to1$ (\Cref{fig:pscan}). The $\mu$-conserving wave loses its passing population faster than the untuned perpendicular wave picks up the trapped one, so the polarization moves the alphas toward the wave that must be strengthened.

\begin{figure}[tbp]
\includegraphics[width=0.95\columnwidth]{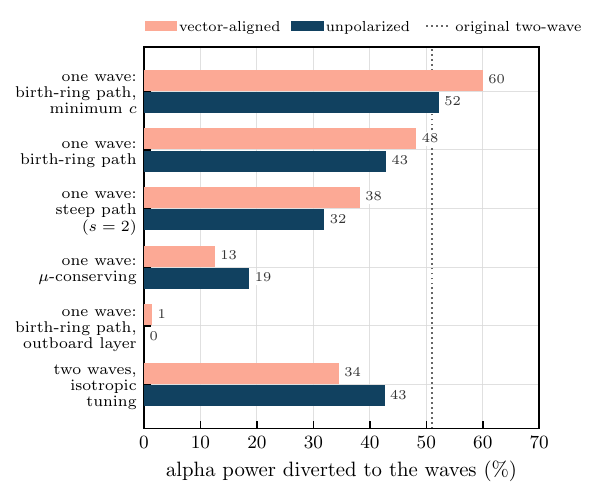}
\caption{Alpha power diverted to the waves for vector-aligned and unpolarized fuel, for the two-wave scheme at its isotropic-fuel tuning and for five single-wave designs, each labeled by its quasilinear path: the birth-ring path $d(\mu B_0)/dE=1$ with the resonance layer at $R_0+0.05\,a$, at $c=1.4$ and at the minimum $c$, the same path with the layer at the original $R_0+0.35\,a$, a $\mu$-conserving path, and a steep path with slope $s=d(\mu B_0)/dE=2$. The dotted line shows the tuned two-wave original~\cite{herrmann1997}.}
\label{fig:design}
\end{figure}

Alignment simplifies the wave design. A wave in this model reaches only the births near its quasilinear path in $(E,\mu B_0)$. Isotropic births cover the whole space, from $\mu B_0=0$ to $E$, and no single path reaches all of them, which is why Herrmann and Fisch needed two waves; vector-aligned births concentrate near $\mu B_0=E$, with banana tips near the axis so one path can reach most of them. Scanning the path slope $d(\mu B_0)/dE$, the flux constant $c$ in $dP_\phi/dE=c\,Ze\,\psi_\mathrm{wall}/\mathcal{E}_0$, and the layer position, the best design has slope $1$, so the cooling population keeps its maximal coupling $\propto\mu B_0/E$; $c=1.02$, near the minimum of one, so each alpha gives the wave nearly all of its energy by the time it reaches the wall; and the layer at $R_0+0.05\,a$, which nearly every orbit crosses. This wave diverts $60\%$ of the alpha power with aligned fuel and $52\%$ with unpolarized (\Cref{fig:design}) compared with $43\%$ for the two-wave scheme at its isotropic tuning and $51\%$ for the tuned original~\cite{herrmann1997}. Its extraction per alpha, nearly the full birth energy at $c=1.02$, is greater than the uniform-field bound $E_\perp$ of \Cref{sec:payoff} because at the banana tip all of a trapped alpha's energy is perpendicular. The other designs of \Cref{fig:design} do worse because the outboard layer misses the core orbits and the $\mu$-conserving path diverts $12\%$ and prefers unpolarized fuel. Restricting the wave to deeply trapped orbits lowers the diverted power to $24\%$ against $16\%$ but raises $A_\eta$ to its pitch-weight bound of $3/2$. We found quadrupling the number of kicks changes nothing, so the diverted fractions are converged rather than limited by the interaction time.

There are a few caveats: First, these are collisionless estimates. Second, we specify each wave by its quasilinear path rather than by a dispersion relation, so we do not check that a real wave would follow that path. Third, the best design puts its resonance layer at $R_0+0.05\,a$, close to the magnetic axis, so a launched wave would have to reach the core relatively undisturbed before it could channel.

\section{Transport model of the ARC plasma}\label{app:transport}

In this appendix we describe our radial transport model for ARC (\Cref{sec:arctransport}) in more detail. We solve for $T_i(\rho)$, $T_e(\rho)$, and $n_\mathrm{He}(\rho)$, with $\rho$ the normalized toroidal-flux label, on the grid of the published V3A profile data~\cite{hillesheim2026data}, resampled to a uniform grid of $101$ points, with the profiles held at the design values near the pedestal top at $\rho=0.9$.

For heat transport, the power through each surface is
\begin{equation}
Q_s=V'\langle|\nabla\rho|^2\rangle n_s T_s \chi_\mathrm{gB}\hat\chi_0\max(z_s-z_{s,\rm crit},0)^{\alpha_\mathrm{stiff}},
\label{eq:Qsflux}
\end{equation}
with $z_s=-d\ln T_s/d\rho$, $\chi_\mathrm{gB}=\rho_i^2 v_{ti}/a$ the gyroBohm diffusivity at the radially local ion temperature, and $\alpha_\mathrm{stiff}=2$ except where we vary it in \Cref{fig:arcmech}(b). We calculate the model's free parameters, the critical-gradient profiles $z_{s,\rm crit}(\rho)$ and the constants $\hat\chi_0$ (for ions and electrons) using power balance. We integrate the design's source profiles to give the power through each surface, $104$ MW in the ion channel and $29$ MW in the electron channel at the edge. We then choose $z_{s,\rm crit}(\rho)$ so that the design profiles conduct this power (according to the flux in \Cref{eq:Qsflux}), and the constant $\hat\chi_0$ so that the mid-radius gradient sits $25\%$ above the critical gradient. The conducted power used in this calibration comes from the source profiles tabulated in the design data; the solver instead recomputes its sources from the models below, which differ from the tabulated ones at the $\sim$percent level. As a test of those models and the discretization, we run the solver from the design profiles with polarization and channeling off: the temperature profiles move by less than $0.3$ keV.

We recompute each heat source from the current profiles at every iteration of the relaxation toward steady state. The alpha power uses the Bosch-Hale reactivity on the local fuel densities, split between ions and electrons by Stix's critical-energy formula~\cite{stix1972}. The radiation has three parts: bremsstrahlung from the local $Z_\mathrm{eff}$, synchrotron scaled from the design profile as $T_e^{5/2}$, and line radiation held fixed. The ion-electron exchange uses the classical $n^2(T_e-T_i)/T_e^{3/2}$ form with its coefficient fitted to the design, applied as an exact, energy-conserving relaxation, and the radio-frequency and ohmic sources are held at the design profiles, $17.7$ MW in total. Benchmarked at the design profiles: the alpha power matches the design to $0.4\%$, the bremsstrahlung to $3\%$, the on-axis ion share is $0.32$ against the design's $0.35$, and the model is self-consistent at $827$ MW of fusion power, a plasma gain of $47$, within the design's own $\sim$$600$ to $1300$ MW range of projections~\cite{hillesheim2026,howard2026}. For the path-deposited variant of \Cref{fig:arcmech}(b), each birth's channeled power is deposited uniformly in $\rho$ from the birth surface to the boundary, conserving power along the outward extraction path described in \Cref{app:com}.

For particle transport we evolve only the helium ash. Its source is the retained birth rate, $(1-F_\mathrm{ej})$ of the alpha births, it diffuses with $D_\mathrm{He}=0.5\,\chi_i$, and its boundary value represents the divertor recycling. In our reference case, which we call the benchmarked helium confinement throughout, the boundary value is $4.8\times10^{18}$ m$^{-3}$, chosen so that the unpolarized basecase has $\tau^*_\mathrm{He}=2.2\,\tau_E$; ejection lowers it by $1-F_\mathrm{ej}$, and the strong-pumping variant sets it near zero. This is better helium exhaust than the $\tau^*_\mathrm{He}/\tau_E\approx4$ demonstrated on JT-60U~\cite{sakasai1999}. On the other hand, it is a conservative choice here because a basecase with lower $\tau^*_\mathrm{He}/\tau_E$ leaves channeling with less helium to remove. Increasing $\tau^*_\mathrm{He}/\tau_E$ so the basecase has $\tau^*_\mathrm{He}=4\,\tau_E$ changes the basecase fusion power by less than $0.5\%$ and leaves the design dilution unchanged since the transfer out of the lumped impurity absorbs the extra ash charge, while the on-axis helium fraction rises from $2.5\%$ to $4.0\%$ and the $\eta_{\rm ch}=0.8$ enhancement rises from $2.2$ to $2.4$.

The design's non-fuel charge is a lumped $Z=5$ impurity that represents the helium ash and any seeded impurities together~\cite{howard2026}, set so the fuel dilution $(n_\mathrm{D}+n_\mathrm{T})/n_e$ is $0.85$. We remove the basecase ash's charge from the lump and return it to the fuel so the basecase keeps the design dilution while the ash density can vary. We enforce quasineutrality by keeping the electron density fixed and displacing one deuteron and one triton for each helium charge, thermal ash and fast alpha, with the fast-alpha density scaled by the local birth rate, the retained fraction $1-F_\mathrm{ej}$, and the slowing-down time. Every enhancement we quote is the ratio of a case's fusion power to that of the unpolarized, unchanneled basecase run with the same helium exhaust assumption.

The magnetic equilibrium is held fixed. The flux-surface volumes, areas, and the averages $\langle|\nabla\rho|\rangle$ and $\langle|\nabla\rho|^2\rangle$ come from contours of the poloidal flux in the published equilibrium, and our model's plasma volume, $184$ m$^3$, matches the volume enclosed by the published boundary to $0.3\%$. Future work should self-consistently recalculate the equilibrium for each set of new profiles.

\section{Alpha-channeling current drive}\label{app:cd}

The waves that channel the alpha power can also drive current, using the power they lose to electrons~\cite{fisch1992}. In the power balance of \Cref{app:wave} we assign the fraction $\zeta/(1+\zeta)$ of the extracted power to electron Landau damping, with $\zeta=0.25$; launched with directed $k_\parallel$, that absorption drives current as in lower-hybrid current drive~\cite{fisch1992,ochsbertelli2015a,ochsbertelli2015b}. At the $\eta_{\rm ch}=0.5$ vector-aligned operating point the alpha power is $p_\alpha\approx1.9$ MW\,m$^{-3}$ and the circulating wave power is $\eta_{\rm ch}(1+\zeta)\,p_\alpha$, so the electrons receive its parasitic share
\begin{equation}
p_e=\zeta\,\eta_{\rm ch}\,p_\alpha\approx0.24\ \mathrm{MW\,m^{-3}}.
\label{eq:pcd}
\end{equation}
Over the plasma volume of the equilibrium we use, $V_p\approx184$ m$^{3}$ (\Cref{app:transport}), this is $P_e\approx44$ MW, and with the lower-hybrid figure of merit $\eta_\mathrm{CD}=I R\, n_{20}/P_e\approx0.3$~\cite{fisch1992}, at the V3A density $n_{20}\approx2.5$ (Greenwald fraction $0.9$~\cite{hillesheim2026}) and $R=4.62$ m,
\begin{equation}
I\approx\frac{\eta_\mathrm{CD} P_e}{R\,n_{20}}\approx1.1\ \mathrm{MA},
\label{eq:Icd}
\end{equation}
about a tenth of the ARC V3A plasma current of $12$ MA~\cite{hillesheim2026}, driven by power otherwise counted as lost, and cheaper for aligned fuel because the inversion lowers the wave's net damping. The near-perpendicular waves computed in this paper have small $k_\parallel$ and drive current poorly per watt absorbed. Realizing this estimate therefore needs a wave spectrum whose parallel phase velocities reach the electrons efficiently~\cite{ochsbertelli2015a,ochsbertelli2015b}.

\clearpage
\bibliography{SPF_alpha_channeling_refs}

\end{document}